\documentstyle[preprint,aps]{revtex}
\begin{document}
\draft
%%%%%%%%%%%% Begin Cover Page %%%%%%%%%%%%%%%%%%%%%%%%%%%%%%%%%%%%%%%%%%
\preprint{ANL-HEP-PR-96-37}
\title{Isolated Prompt Photon Production in Hadronic\\ 
       Final States of $e^+e^-$ Annihilation }
\author{Edmond L. Berger$^1$, Xiaofeng Guo$^2$, and Jianwei Qiu$^2$}
\address{$^1$High Energy Physics Division,
             Argonne National Laboratory \\
             Argonne, Illinois 60439, USA \\
         $^2$Department of Physics and Astronomy, 
             Iowa State University \\
             Ames, Iowa 50011, USA }
\date{May 15, 1996}
\maketitle

\begin{abstract} 
We provide complete analytic expressions for the isolated prompt
photon production cross section in $e^+e^-$ annihilation reactions
through one-loop order in quantum chromodynamics (QCD) perturbation
theory.  Functional dependences on the isolation cone size $\delta$
and isolation energy parameter $\epsilon$ are derived.  The energy
dependence as well as the full angular dependence of the cross section
on $\theta_\gamma$ are
displayed, where $\theta_\gamma$ specifies the direction  of the
photon with respect to the $e^+e^-$ collision axis.  We point out that
conventional perturbative QCD factorization breaks down for
isolated photon production in $e^+e^-$ annihilation reactions  
in a specific region of phase space. We discuss the implications of
this breakdown for the extraction of fragmentation functions from
$e^+e^-$ annihilation data and for computations of prompt photon
production in hadron-hadron reactions. 
\end{abstract} 
\vspace{0.2in}
%%%%%%%%%%%% End of Cover Page %%%%%%%%%%%%%%%%%%%%%%%%%%%%%%%%%%%%%%%%%

%%%%%%%%%%%%%% Begin Section I %%%%%%%%%%%%%%%%%%%%%%%%%%%%%%%%%%%%%%%%%
\section{Introduction}
\label{sec:1}

The cross section for the inclusive yield of high energy photons 
in hadronic final states of $e^+e^-$ annihilation is well-defined
and, as demonstrated in our earlier paper \cite{BGQ1}, it may be
calculated reliably within the context of quantum chromodynamics (QCD)
perturbation theory.  However, 
an important practical limitation of high energy investigations, 
whether in hadron-hadron reactions or $e^+e^-$ annihilation processes, 
is that photons are observed and their cross sections are measured
only when the photons are relatively isolated, separated to 
some extent in phase space from accompanying hadrons.  Experimental
procedures vary, but the essence of isolation is that a cone of
half-angle $\delta$ is 
drawn about the direction of the photon's momentum, as sketched in
Fig.~\ref{fig0}, and the cross section is defined for photons 
accompanied by less than a specified amount of hadronic energy in the
cone, {\it e.g.}, $E_h^{cone} \leq E_{\max}\equiv \epsilon_h
E_\gamma$.  In this paper, we undertake a thorough analysis of
isolated prompt photon production in $e^+e^-$ annihilation.  Previous
theoretical studies of isolated prompt photon production in
$e^+e^- \rightarrow \gamma X$ include those of 
Refs.~\cite{EWNG,KT,BGQ,BGQ2}.

In this paper, our calculations of photon yields in
$e^+e^- \rightarrow \gamma X$ are carried out through one-loop order.
We compute explicitly direct photon production through first order in
the electromagnetic coupling strength, $\alpha_{em}$, and the
quark-to-photon and gluon-to-photon fragmentation contributions
through first order in the strong coupling strength $\alpha_s$.  We
display the full angular dependence of the cross sections, separated
into longitudinal $\sin^2 \theta_\gamma$ and
transverse components $\left( 1 + \cos^2 \theta_\gamma \right)$,
where $\theta_\gamma$ is the direction of the $\gamma$ with respect to
the $e^+e^-$ collision axis.

A proper theoretical treatment of the isolated energetic photon yield
requires careful consideration of the origins of both infrared and
collinear singularities in QCD perturbation theory.  In a
theoretical calculation, photon isolation limits the final-state phase
space accessible to accompanying gluons (g) and quarks (q).  This
phase space restriction breaks the perfect cancellation of soft
singularities in each order of perturbation theory that guarantees
reliable predictions in the inclusive case \cite{BGQ2}.  An
uncanceled infrared singularity signals a breakdown of conventional
perturbative factorization.  In this case, the uncanceled infrared 
singularity leads to an inverse-power divergence at the
level of the partonic cross section [i.e., a $1/(1-x_1)$ divergence as
$x_1\rightarrow 1$] in the order $\alpha_s$ one-loop quark-to-photon 
fragmentation term.  After convolution with the
parton-to-photon fragmentation function, this inverse-power divergence
becomes a logarithmic divergence in the isolated cross section.  

Owing to isolation, the predicted photon cross section
develops explicit functional dependence on the isolation parameters 
$\epsilon_h$ and $\delta$.  We show that at one-loop level, this cross
section is well defined in most of phase space, so long as
$\epsilon_h$ and $\delta$ are not too small.  However, the region near
$x_\gamma=1/(1+\epsilon_h)$ is special, and the 
perturbatively calculated cross section is
singular there; $x_\gamma = 2 E_\gamma/\sqrt s$, where $\sqrt s$
denotes the center-of-mass energy of the $e^+e^-$ reaction.  
In earlier papers \cite{BQ}, we presented analytic expressions for the
dependence on isolation parameters in the specific case of photon
production in hadron-hadron collisions \cite{Baer,Aurenche,GV}.  In
this paper, 
we provide a much more detailed treatment of photon isolation, 
concentrating on energetic photon production in electron-positron
annihilation: $e^+e^- \rightarrow \gamma X$ \cite{BGQ,BGQ2}.  
 
In a perturbative QCD (pQCD) calculation of the inclusive yield 
of photons, the quark-photon collinear singularities that arise in
each order of perturbation theory, associated with the hadronic
component of the photon, are subtracted and absorbed into
quark-to-photon and gluon-to-photon fragmentation functions,
$D(z,\mu^2)$, in accord with the factorization theorem \cite{AEMP}.
The scale $\mu^2$ denotes the fragmentation scale that separates the
non-perturbative domain from the region in which perturbation theory
should apply; $z$ is the momentum fraction carried by the observed
photon from its parent parton.  Since fragmentation is a process in
which photons are part of quark, antiquark, or gluon ``jets", it is
evident that photon isolation reduces the contribution from
fragmentation terms.  For a small value of the energy resolution
parameter, $\epsilon_h$, the isolation cut eliminates most of the
contribution from parton-to-photon fragmentation.  However,
$\epsilon_h$ may never be equal to zero either experimentally or
theoretically.  Experimentally, the finiteness of $\epsilon_h$ is 
guaranteed by detector resolution.  Theoretically, the perturbative 
calculation of the cross section for isolated photons is ill  
defined if $\epsilon_h$ vanishes \cite{BQ}.  

Fragmentation is modeled in perturbation theory as a collinear
process, whereas experimental cone sizes are finite $(\delta \neq 0)$
and partons are manifested as sprays of hadrons.  Correspondingly,
there is an inherent conceptual incompatibility between theoretical,
collinear fragmentation functions and empirical fragmentation
functions, $D_{\exp} \left(z,\mu^2,\delta\right)$, that more naturally
would be defined with reference to a cone of specified size.  In this
paper, we limit ourselves to the usual collinear fragmentation  
functions, $D_{c\rightarrow\gamma}\left(z,\mu^2\right)$. 
We derive the expected dependence of the cross section on cone size 
$\delta$ and energy isolation $\epsilon_h$, and we display a limited 
region of phase space in which the incompatibility of the collinear 
and finite cone size assumptions leads to collinear
sensitivity of the isolated photon cross section.  We discuss the
impact of both infrared and collinear sensitivity on computations of
prompt photon production in electron-positron and hadron-hadron
reactions.    

All four groups at LEP have published
papers on prompt photon production \cite{LEP4}.  A measurement of the
photon fragmentation function from an analysis of the two-jet rate in
Z decays is reported by the ALEPH
collaboration \cite{LEP5}.  Isolated prompt photon production has
been treated theoretically previously, but our approach is different.
In Ref.~\cite{EWNG}, the authors concentrate on events having a
specific exclusive topology, such as a photon plus one hadronic jet;
they discuss the extraction of the quark to photon fragmentation
function from such data.  In Ref.~\cite{KT}, isolated direct photon
production through order $\alpha_{em}$ is treated; the fragmentation
terms there are included only at lowest order, not through order
$\alpha_s$, and the angular dependence of the cross section is not
derived.  Practical aspects of confronting theoretical calculations
with data from LEP are addressed in Ref.~\cite{LEP}.   

In Section~\ref{sec:2}, we discuss the similarities and
differences in the theoretical calculations of the cross sections for 
inclusive and isolated photons, and we provide a general factorized
formula for the isolated photon cross section.  Based on the
definition of isolation, it is clear that the isolated photon cross
section is part of the inclusive cross section.  We write the isolated
cross section as the inclusive cross section minus a subtraction term
\cite{BQ}.  As in the calculation of jet cross sections \cite{SW}, the
singularity structure of subtraction term is much easier to deal with
than that of the isolated cross section itself, at the order in
perturbation theory in which we are working.  To establish notation,
we derive explicit expressions for the isolated photon yield in lowest
order $\left( O\left( \alpha^o_{em}\right),\ O\left(\alpha^o_s\right)
\right)$.  In Section~\ref{sec:3}, we present our derivation 
of the three one-loop contributions to the subtraction terms.
We work in $n = 4-2\epsilon$ dimensions in order to display
singularities explicitly.  The hard-scattering matrix elements are
identical to those for the fully inclusive case discussed in
Ref.~\cite{BGQ1}, but the integration over momentum variables of the
unobserved final-state partons is restricted by the requirements of
photon isolation.  All of our calculations are done analytically.  In
Section IV, we summarize our final expressions for the isolated photon
cross section $E_\gamma d\sigma ^{iso}_{e^+e^- \rightarrow \gamma
X}/d^3\ell$, and we discuss the logarithmic divergence,
$\ell n|(1+\epsilon_h)-1/x_\gamma|$, associated with the 
limiting case in which $x_\gamma$ approaches $1/(1+\epsilon_h)$.  
Numerical estimates, implications for isolated prompt photon
phenomenology in $e^+e^-$ and hadron-hadron reactions, thoughts for
further work, and suggestions for comparisons with data are collected
in Section V.   
%%%%%%%%%%%%%% End of Section I %%%%%%%%%%%%%%%%%%%%%%%%%%%%%%%%%%%%%%%%%

%%%%%%%%%%%%%% Begin Section II %%%%%%%%%%%%%%%%%%%%%%%%%%%%%%%%%%%%%%%%
\section{Factorization and Lowest Order Contribution}
\label{sec:2}

In this section we drive a general factorized formula for
the cross section for isolated photons, and we end the section with
a computation of the lowest order 
$O\left(\alpha^o_{em}\alpha^o_s\right)$ contribution to the isolated
energetic photon yield in $e^+e^- \rightarrow \gamma X$.  The general
notation in this paper is the same as that in Ref.~\cite{BGQ1}.

\subsection{Factorized Form for the Isolated Cross Section}
\label{subsec:2a}

With the assumption of parton-to-photon factorization, the cross
section for an $m$ parton final state in $e^+e^-\rightarrow \gamma X$,
as sketched in Fig.~\ref{fig1}, is 
\begin{equation}
d\sigma^{(m)} = {{1} \over {2s}}\Big| \overline{M}
_{e^+e^-\rightarrow {\underbrace{c+\cdots}_{m}}}
\Big|^2 dPS^{(m)}\cdot dz
D_{c\rightarrow \gamma} (z).
\label{e1}
\end{equation}
Parton $c = \gamma, q, \bar{q}, g$, and $z = E_\gamma/E_c$.
The factor $dPS^{(m)}$ is the $m$-parton phase space, and 
$D_{c\rightarrow\gamma}(z)$ is a function that describes fragmentation
of parton $c$ 
to the photon.  This general expression is valid for both inclusive
and isolated cross sections.  The difference between the inclusive and
isolated cross sections resides in the phase space integration for the
final state partons.  For isolated cross sections, because of the
isolation condition, not all partons can be integrated over all phase
space.  For example, partons with energy larger than $\epsilon_h
E_{\gamma}$ are excluded from the cone of isolation about 
the observed photon.

As we proposed in Ref.~\cite{BQ}, we write
the isolated cross section as the following difference of cross 
sections:
\begin{equation}
E_\gamma{{ d\sigma^{iso}_{e^+e^- \rightarrow \gamma X}} \over 
{d^3\ell}} = 
E_\gamma {{d\sigma^{incl}_{e^+e^- \rightarrow \gamma X}} \over
{d^3\ell}} - 
E_\gamma{{ d\sigma^{sub}_{e^+e^- \rightarrow \gamma X}} \over 
{d^3\ell} }.
\label{e2}
\end{equation}
In Eq.~(\ref{e2}), $E_\gamma d\sigma^{incl}/d^3\ell$ is 
the cross section for inclusive photons.  It is 
well-defined in QCD perturbation theory and can be expressed in 
the following factorized form \cite{BGQ1}
\begin{eqnarray}
E_\gamma \frac{d\sigma^{incl}_{e^+e^- \rightarrow \gamma X}}{d^3\ell}
&=& \sum_c\  
    \int_{x_\gamma}^1 \frac{dz}{z^2}\
    E_c \frac{d\hat{\sigma}^{incl}_{e^+e^- \rightarrow cX}}{d^3 p_c} 
    \left(x_c=\frac{x_\gamma}{z}\right) 
    D_{c \rightarrow \gamma}(z)
    \nonumber \\
&\equiv & \sum_c\  
    E_c \frac{d\hat{\sigma}^{incl}_{e^+e^- \rightarrow cX}}{d^3 p_c}
    \otimes D_{c \rightarrow \gamma}(z)\ ,
\label{e3}
\end{eqnarray}
where $x_c=2E_c/\sqrt{s}$, and the sum  
extends over $c = \gamma,q,\bar{q}$ and $g$. 
The short-distance hard-scattering cross sections 
$E_c d\hat{\sigma}^{incl}_{e^+e^- \rightarrow cX}/d^3 p_c$ 
in Eq.~(\ref{e3}) are derived in  
Ref.~\cite{BGQ1} through one-loop level.  

Cross sections for isolated photons 
$E_\gamma d{\sigma}^{iso}_{e^+e^- \rightarrow \gamma X}/d^3\ell$ 
measured, e.g., in LEP experiments, are well-behaved and finite.
Therefore, the theoretical subtraction term 
$E_\gamma d{\sigma}^{sub}_{e^+e^- \rightarrow \gamma X}/d^3\ell$,
defined in Eq.~(\ref{e2}), should be well-behaved as well.
Since the available phase space for the isolated photon cross section
is smaller than that for inclusive photons, the cross section for
isolated photons should be less than the corresponding inclusive cross
section, 
\begin{equation}
E_\gamma \frac{d\sigma^{iso}_{e^+e^- \rightarrow \gamma X}}{d^3\ell}
\leq
E_\gamma \frac{d\sigma^{incl}_{e^+e^- \rightarrow \gamma X}}{d^3\ell} 
\ .
\label{e3p}
\end{equation}
Consequently, the subtraction term, 
$E_\gamma d{\sigma}^{sub}_{e^+e^- \rightarrow \gamma X}/d^3\ell$,
should be positive and finite.  In terms of the definition given in
Eq.~(\ref{e2}), 
$E_\gamma d{\sigma}^{sub}_{e^+e^- \rightarrow \gamma X}/d^3\ell$
can be viewed as a ``cross section'' for a photon ``jet'' with 
photon momentum $\ell$ and hadronic energy 
in the ``jet'' cone $E_h^{cone}$ restricted to be 
\underline{larger} than $E_{\rm max} = \epsilon_h E_\gamma$.  
The virtue of Eq.~(\ref{e2}) is that the infrared and collinear
singularities of the subtracted term $\sigma^{sub}$
are much easier to deal with, at the order in which we are working, 
than those of $\sigma^{iso}$ itself.

We \underline{assume} as a working hypothesis that the subtraction 
term $\sigma^{sub}_{e^+e^- \rightarrow \gamma X}$ 
can be factored in the same way as the inclusive cross section and 
expressed as a convolution:
\begin{equation}
E_\gamma \frac{d\sigma^{sub}_{e^+e^- \rightarrow \gamma X}}{d^3\ell}
= \sum_c  
    \int_{x_\gamma}^1 \frac{dz}{z^2}\
    E_c \frac{d\hat{\sigma}^{sub}_{e^+e^- \rightarrow cX}}{d^3 p_c} 
    \left(x_c=\frac{x_\gamma}{z}\right) 
    D_{c \rightarrow \gamma}^{iso} (z,\delta)\ .
\label{e4}
\end{equation}
In Eq.~(\ref{e4}), we include explicit dependence on the cone size
$\delta$ in the fragmentation functions in order to point out that, in
principle, the fragmentation functions extracted from the cross
section for isolated photons may depend on the definition of the
isolation cone.  Since the inclusive cross section $\sigma^{incl}$ in
Eq.~(\ref{e2}) is well-defined in QCD perturbation theory, it is our
principal task in this paper to show to what extent the short-distance
subtraction terms $\hat{\sigma}^{sub}_{e^+e^-\rightarrow cX}$ are free 
from infrared and collinear divergences.  As we show in Sec.~IV,
the conventional factorization Eq.~(\ref{e4}) fails in the
neighborhood of a well defined point in phase space,
$x_\gamma=2E_\gamma/\sqrt{s} = 1/(1+\epsilon_h)$. 

The limits of integration over $z$ in Eq.~(\ref{e4}) are fixed by
kinematics: $x_c=x_\gamma/z \leq 1$.  However, the isolation 
condition imposes, in addition, a requirement on the total hadronic
energy in the isolation cone, $E_h^{cone} \geq E_{\max}=\epsilon_h
E_\gamma$.  Generally, $E_h^{cone}$ has two sources: 
$E_{frag} + E_{partons}$, as is illustrated in Fig.~\ref{fig2}.  
The fragmentation component, $E_{frag}$, is the hadronic component 
of the ``jet'' from which the energetic photon itself emerges.
In the collinear approximation,
\begin{equation}
E_{frag} = (1-z)E_c = \left(\frac{1-z}{z}\right)E_\gamma.
\label{e5}
\end{equation}
The second component, $E_{partons}$, is contributed by \underline{other} 
final state partons that are emitted into the region of phase space 
defined by the photon isolation cone.  For the subtraction term,
\begin{eqnarray}
E^{cone}_h 
&=& E_{partons} + E_{frag} \nonumber \\
&=& E_{partons} + \left( \frac{1-z}{z} \right) 
E_\gamma \geq E_{\max} \equiv \epsilon_h E_\gamma .
\label{e6}
\end{eqnarray}

If $z \leq 1/(1 + \epsilon_h)$, or, equivalently,
$E_{frag} \geq \epsilon_h E_\gamma$, the constraint 
$E_h^{cone} \geq \epsilon_h E_\gamma$ is satisfied for any value of 
$E_{partons}$.  Correspondingly, there is no 
restriction on the phase space for accompanying final state partons.
Consequently, 
$\hat{\sigma}^{sub}_{e^+e^- \rightarrow cX} = 
\hat{\sigma}^{incl}_{e^+e^- \rightarrow cX}$ 
for $z \leq 1/(1 + \epsilon_h)$.

If $z > 1/(1 + \epsilon_h)$, to satisfy 
$E_h^{cone} \geq \epsilon_h E_\gamma$, it is necessary that 
\begin{equation}
E_{partons} \geq E_{\max} - E_{frag}
\equiv E_{\min}(z) 
= \left[ (1 + \epsilon_h) - \frac{1}{z}\right] E_\gamma.
\label{e7}
\end{equation}
We remark that $E_{\min} (z) > 0$ as long as 
$z > 1/(1 + \epsilon_h)$, and $E_{\min} (z) = 0$ if
$z = 1/(1 + \epsilon_h)$.

Taking into account the constraints from the isolation condition, we can 
rewrite Eq.~(\ref{e4}) as 
\begin{eqnarray}
E_\gamma \frac{d\sigma^{sub}_{e^+e^- \rightarrow \gamma X}}{d^3\ell}
&=& \sum_c
    \int_{\max\left[x_\gamma,\frac{1}{1+\epsilon_h}\right]}^1 
    \frac{dz}{z^2}\
    E_c \frac{d\hat{\sigma}^{sub}_{e^+e^- \rightarrow cX}}{d^3 p_c} 
    \left(x_c=\frac{x_\gamma}{z}\right) 
    \Bigg|_{E_{partons}\geq E_{\min}}
    D^{iso}_{c \rightarrow \gamma}(z,\delta)
    \nonumber \\
&+& \sum_c
    \int_{x_\gamma}^{\max\left[x_\gamma,\frac{1}{1+\epsilon_h}\right]} 
    \frac{dz}{z^2}\
    E_c \frac{d\hat{\sigma}^{incl}_{e^+e^- \rightarrow cX}}{d^3 p_c} 
    \left(x_c=\frac{x_\gamma}{z}\right) 
    D^{iso}_{c \rightarrow \gamma}(z,\delta) 
\label{e8} \\
&\equiv&
    \sum_c \left[
    E_c \frac{d\hat{\sigma}^{sub}_{e^+e^- \rightarrow cX}}{d^3 p_c} 
    \dot{\otimes} 
    D^{iso}_{c \rightarrow \gamma}(z,\delta)
 +  E_c \frac{d\hat{\sigma}^{incl}_{e^+e^- \rightarrow cX}}{d^3 p_c} 
    \ddot{\otimes}  
    D^{iso}_{c \rightarrow \gamma}(z,\delta)
    \right]\ . \nonumber
\end{eqnarray}
It is important to note in Eq.~(\ref{e8}) that  
the short-distance subtraction terms 
$E_c d\hat{\sigma}^{sub}_{e^+e^- \rightarrow cX}/d^3 p_c$ 
are needed only 
for $x_c\leq {\min}[x_\gamma (1+\epsilon_h),1] < 1$, if $x_\gamma <
1/(1+\epsilon_h)$.

The modified convolution signs ``$\dot{\otimes}$'' and 
``$\ddot{\otimes}$'' in Eq.~(\ref{e8}) are defined in the same way as 
the convolution sign ``$\otimes$'' in Eq.(\ref{e3}), except for the limits 
of the $z$-integration.  For functions $A(x_c)$ and $B(z)$, we define
\begin{mathletters}
\label{e8t}
\begin{eqnarray}
A(x_c)\otimes B(z) \equiv \int_{x_\gamma}^1\ \frac{dz}{z^2}\ 
A(x_c=x_\gamma/z)\ B(z)\ ; 
\label{e8a} \\
A(x_c)\dot{\otimes} B(z) \equiv 
\int_{\max\left[x_\gamma,\frac{1}{1+\epsilon_h}\right]}^1\
\frac{dz}{z^2}\ A(x_c=x_\gamma/z)\ B(z)\ ; 
\label{e8b}  \\
A(x_c)\ddot{\otimes} B(z) \equiv 
\int_{x_\gamma}^{\max\left[x_\gamma,\frac{1}{1+\epsilon_h}\right]}\ 
\frac{dz}{z^2}\ A(x_c=x_\gamma/z)\ B(z)\ .
\label{e8c}
\end{eqnarray}
\end{mathletters}
Notice the identity $\otimes = \dot{\otimes} + \ddot{\otimes}$.  

Substituting Eqs.~(\ref{e3}) and (\ref{e8}) into Eq.~(\ref{e2}), we 
derive a simplified expression for the isolated cross section
\begin{eqnarray}
E_\gamma \frac{d\sigma^{iso}_{e^+e^- \rightarrow \gamma X}}{d^3\ell}
&=& \sum_c\  
    \int_{\max\left[x_\gamma,\frac{1}{1+\epsilon_h}\right]}^1 
    \frac{dz}{z^2}\left(
    E_c \frac{d\hat{\sigma}^{incl}_{e^+e^- \rightarrow cX}}{d^3 p_c} 
   -E_c \frac{d\hat{\sigma}^{sub}_{e^+e^- \rightarrow cX}}{d^3 p_c} 
    \Bigg|_{E_{partons}\geq E_{\min}}
    \right) D_{c \rightarrow \gamma}(z)
    \nonumber \\
&=& \sum_c\left(
    E_c \frac{d\hat{\sigma}^{incl}_{e^+e^- \rightarrow cX}}{d^3 p_c}
   -E_c \frac{d\hat{\sigma}^{sub}_{e^+e^- \rightarrow cX}}{d^3 p_c}
    \right) \dot{\otimes} D_{c \rightarrow \gamma}(z)\ .
\label{e9a}
\end{eqnarray} 
In deriving Eq.~(\ref{e9a}), we assume for simplicity  
$D_{c \rightarrow \gamma}^{iso} (z,\delta) 
= D_{c \rightarrow \gamma}(z)$.

When $c = \gamma,\ D_{c\rightarrow \gamma} (z) = \delta(1-z)$ through
order $O(\alpha_{em})$, and $E_{frag} = 0$ (i.e., 
$E_{\min} = \epsilon_h E_\gamma)$. 
Therefore, we may rewrite Eq.~(\ref{e9a}) more explicitly as
\begin{equation}
E_\gamma \frac{d\sigma^{iso}_{e^+e^- \rightarrow \gamma X}}
              {d^3\ell} 
= 
E_\gamma \frac{d\hat{\sigma}^{iso}_{e^+e^-\rightarrow\gamma X}}
              {d^3\ell} 
+ \sum_{c = q,\bar{q},g}\, 
E_c\frac{d\hat{\sigma}^{iso}_{e^+e^- \rightarrow cX}}{d^3 p_c}\,
\dot{\otimes}\, D_{c\rightarrow \gamma} (z)\ .
\label{e9}
\end{equation}
The isolated short-distance hard-scattering cross sections 
are defined as
\begin{mathletters}
\label{e10}
\begin{eqnarray}
E_\gamma \frac{d\hat{\sigma}^{iso}_{e^+e^-\rightarrow\gamma X}}
              {d^3\ell} 
&\equiv &
E_\gamma \frac{d\hat{\sigma}^{incl}_{e^+e^-\rightarrow\gamma X}}
              {d^3\ell} - 
E_\gamma \frac{d\hat{\sigma}^{sub}_{e^+e^-\rightarrow\gamma X}}
              {d^3\ell}
         \Bigg|_{\stackrel{partons\ inside\ cone;}
                {E_{partons \geq \epsilon_h E_\gamma.}}}  
\label{e10a} \\
E_c \frac{d\hat{\sigma}^{iso}_{e^+e^- \rightarrow cX}}{d^3 p_c} 
&\equiv &
  E_c \frac{d\hat{\sigma}^{incl}_{e^+e^- \rightarrow cX}}{d^3 p_c} 
- E_c \frac{d\hat{\sigma}^{sub}_{e^+e^- \rightarrow cX}}{d^3 p_c} 
\Bigg|_{\stackrel{partons\ inside\ cone;}{E_{partons \geq E_{\min}.}}} .
\label{e10b}
\end{eqnarray}
\end{mathletters}
The value of $E_{\min}$ is specified in Eq.~(\ref{e7}).  

Anticipating a result to be derived below, we remark that when
$x_\gamma \rightarrow 1/(1+\epsilon_h)$, 
the subtraction term $\hat{\sigma}^{sub}$ and, consequently, the 
isolated cross section $\hat{\sigma}^{iso}$ develop a logarithmic 
singularity, $\ell n |(1+\epsilon_h)-1/x_\gamma|$.  
We discuss this singularity in detail in Section~\ref{subsec:4d}.

\subsection{Derivation of the Lowest Order Contribution}
\label{subsec:2b}

In \underline{lowest}\ \underline{order}, 
$O\left( \alpha^o_{em} \alpha^o_s \right)$, photon production occurs
only through the quark or antiquark fragmentation process sketched in  
Fig.~\ref{fig3}.  In this case $c = q, \bar{q}$, and
$x_c = 2E_c/\sqrt{s} = 1$ in Eq.~(\ref{e9a});
and there is no direct production of photons.  
Owing to momentum balance, the quark and anti-quark have equal but
opposite momentum in the overall center-of-mass system, and there can
be \underline{no} accompanying parton in the isolation cone around the
fragmenting parton, i.e., $E_{partons} = 0$ inside the cone.
Therefore, at $O\left( \alpha^0_{em} \alpha^0_s \right)$,
\begin{equation}
E_c \frac{d\hat{\sigma}^{(0)sub}_{e^+e^-\rightarrow cX}}
         {d^3p_c}
\Bigg|_{\stackrel{partons\ inside\ cone;}{E_{partons \geq E_{\min}.}}}
\ = 0\ .
\label{e11a}
\end{equation}
Substituting Eq.~(\ref{e11a}) into Eq.~(\ref{e9a}), we derive
\begin{equation}
E_\gamma \frac{d\sigma^{(0)iso}_{e^+e^-\rightarrow \gamma X}}{d^3\ell}
= \int^1_{\max\left[ x_\gamma, \frac{1}{1+\epsilon_h}\right]}
\frac{dz}{z^2}\, \sum_{c=q,\bar{q}} 
E_c \frac{d\hat{\sigma}^{(0)incl}_{e^+e^-\rightarrow cX}}{d^3p_c}
\left( x_c = {{x_\gamma} \over {z}}\right) D_{c\rightarrow \gamma} (z).
\label{e11}
\end{equation}
The lowest order hard-scattering cross section
$E_c d\hat{\sigma}^{(0)incl}_{e^+e^-\rightarrow cX}/d^3p_c$ 
in $n$ dimensions is derived in our earlier paper \cite{BGQ1}:
\begin{equation}
E_c \frac{d\hat{\sigma}^{(0)incl}_{e^+e^-\rightarrow cX}}{d^3 p_c} =
\left[\frac{2}{s} F^{PC}_c (s)\right]\, \alpha^2_{em} N_c 
\left( \frac{4\pi\mu^2}{(s/4)\sin^2\theta_c}\right)^\epsilon 
\frac{1}{\Gamma(1-\epsilon)}
\left[ \left( 1 + \cos^2\theta_c \right) -2\epsilon \right]
\frac{\delta(x_c-1)}{x_c},
\label{e12}
\end{equation}
with $\epsilon=(4-n)/2$, and $c=q,\bar{q}$.
The angle $\theta_c$ is the scattering angle of parton $c$ with
respect to the $e^+e^-$ collision axis in the overall center-of-mass
frame.  The normalization factor is $(2/s)F^{PC}_q(s)$ is \cite{BGQ1}
\begin{eqnarray}
\frac{2}{s} F^{PC}_{q}(s)=\frac{1}{s^{2}} &\Bigg[
& e_{q}^{2}
+ \left(|v_e|^2 + |a_e|^2\right) \left(|v_q|^2 + |a_q|^2\right)
  \frac{s^2}{\left( s-M^2_Z\right)^2 + M^2_Z \Gamma^2_Z} \nonumber \\
&-&  2 e_{q}v_{e}v_{q} 
  \frac{s \left( s-M_{Z}^{2} \right)}
       {\left( s-M^2_Z\right)^2 + M^2_Z \Gamma^2_Z}\, \Bigg] \ .
\label{e12a}
\end{eqnarray}
In Eq.~(\ref{e12a}), both $\gamma$ and $Z^\circ$ intermediate state
contributions are represented, including their interference.  The 
vector $(v)$ and axial-vector $(a)$ 
couplings are defined through $ie\gamma^\mu(v_f+a_f\gamma_5)$, 
the vertex coupling between the intermediate vector boson and the 
initial/final fermion pair of flavor $f$ \cite{BGQ1}.

At this order, the cross section is manifestly finite in the limit 
$\epsilon \rightarrow 0$, and we may set $\epsilon = 0$ directly in 
Eq.~(\ref{e12}).  Nevertheless, Eq.~(\ref{e12}) expressed in 
$n$ dimensions is valuable for later comparison with the 
higher order cross section.
Substituting Eq.~(\ref{e12}) into Eq.~(\ref{e11}), 
we obtain the lowest order isolated cross section \cite{BGQ}
\begin{equation}
E_\gamma {{d\sigma^{(0)iso}_{e^+e^-\rightarrow \gamma X}} 
  \over d^3 \ell}
= 2\sum_q \left[ {{2} \over {s}} F^{PC}_q (s)\right] \alpha^2_{em}
N_c (1 + \cos^2\theta_\gamma) {{1} \over {x_\gamma}}
D_{q \rightarrow \gamma} (x_\gamma, \mu_F),
\label{e13}
\end{equation}
for $x_\gamma \geq 1/(1+ \epsilon_h)$.  At this order, the isolated
cross section vanishes if $x_\gamma < 1/(1+ \epsilon_h)$.  The angles
$\theta_\gamma$ and $\theta_c$ are identical since we take all
products of the fragmentation to be collinear.  The overall factor of
2 in Eq.~(\ref{e13}) accounts for the $\bar{q}$ contribution.  Our
result in Eq.~(\ref{e13}) is consistent with those derived in
Refs.~\cite{EWNG,KT}.
%%%%%%%%%%%%%% End of Section II %%%%%%%%%%%%%%%%%%%%%%%%%%%%%%%%%%%%%%%

%%%%%%%%%%%%%% Begin Section III %%%%%%%%%%%%%%%%%%%%%%%%%%%%%%%%%%%%%%
\section{Subtraction Terms at One-Loop Order}
\label{sec:3}

Having calculated the cross section for inclusive photon production 
in Ref.~\cite{BGQ1}, we must calculate the subtraction 
terms defined in Eqs.~(\ref{e2}) and (\ref{e10}) 
in order to derive the complete 
expression for the isolated photon cross section.  
As in the inclusive case, there are three distinct contributions 
to the subtraction terms for $e^+e^-\rightarrow \gamma X$ 
at one-loop level in perturbation theory:
\begin{mathletters}
\label{eq14}
\begin{eqnarray}
e^+e^- \rightarrow \gamma, &\hspace{0.5in}&
O(\alpha_{em}) \label{e14a}\\
e^+e^- \rightarrow q\ ({\rm or}\ \bar{q}) \rightarrow \gamma, 
&\hspace{0.5in}& O(\alpha_s)
\label{e14b}\\
&\rm{and}& \nonumber \\
e^+e^- \rightarrow g \rightarrow \gamma. 
&\hspace{0.5in}& O(\alpha_s)
\label{e14c}
\end{eqnarray}
\end{mathletters}                      
In Eqs. (\ref{e14b}) and (\ref{e14c}), we have in mind contributions 
from quark (or antiquark) and gluon
fragmentation to photons in the three-parton final state process 
$e^+e^- \rightarrow q\bar{q}g$.  The first contribution,
Eq.~(\ref{e14a}), arises from $e^+e^- \rightarrow q\bar{q}\gamma$
where the $\gamma$ is \underline{not} collinear with either $\bar{q}$
or $q$. 

In this section we derive and present explicit expressions for 
the contributions to 
$E_\gamma d\sigma^{sub}_{e^+e^- \rightarrow\gamma X}/d^3\ell$ 
from each of the three processes in Eq.~(\ref{eq14}).

\subsection{Factorized Form for the
Short-Distance Hard Parts: $\hat{\sigma}^{sub}$}
\label{subsec:3a}

To calculate the short-distance partonic cross sections (hard parts), 
$\hat{\sigma}^{sub}$ in Eq.~(\ref{e10}), we first apply the 
factorized form in Eq.~(\ref{e8}) to parton states 
perturbatively order-by-order in the coupling constants, and
we then extract the perturbative expressions for 
$\hat{\sigma}^{sub}$.  We illustrate this procedure in this subsection
for each of the processes in Eq.~(\ref{eq14})

The Feynman graphs for $\sigma^{(1)sub}_{e^+e^- \rightarrow\gamma X}$
are sketched in Fig.~\ref{fig4}.  To derive the direct production 
term corresponding to Eq.~(\ref{e14a}),
$\hat{\sigma}^{(1)sub}_{e^+e^- \rightarrow \gamma X}$, 
we apply Eq.~(\ref{e8}) perturbatively 
to first order in $\alpha_{em}$.  We obtain
\begin{eqnarray}
\left. \sigma^{(1)sub}_{e^+e^- \rightarrow\gamma X} 
   \right|_{E_q (or E_{\bar{q}})\geq \epsilon_h E_\gamma}
&=&\hat{\sigma}^{(1)sub}_{e^+e^- \rightarrow\gamma X} (x_\gamma)
   \dot{\otimes} D^{(0)}_{\gamma\rightarrow\gamma} (z)
 + \hat{\sigma}^{(1)incl}_{e^+e^- \rightarrow\gamma X} (x_\gamma)
   \ddot{\otimes} D^{(0)}_{\gamma\rightarrow\gamma} (z)
\nonumber \\
&+&\hat{\sigma}^{(0)sub}_{e^+e^- \rightarrow qX} (x_q)
   \dot{\otimes} D^{(1)}_{q\rightarrow\gamma} (z)
 + \hat{\sigma}^{(0)incl}_{e^+e^- \rightarrow q X} (x_q)
   \ddot{\otimes} D^{(1)}_{q\rightarrow \gamma} (z)
\nonumber \\
&+&(q \rightarrow \bar{q})\ .
\label{e15}
\end{eqnarray}
The zeroth order subtraction term 
$\hat{\sigma}^{(0)sub}_{e^+e^- \rightarrow cX}$ vanishes,
as shown in Eq.~(\ref{e11a}).  Since the zeroth order photon-photon  
fragmentation function $D^{(0)}_{\gamma\rightarrow\gamma}(z) =
\delta(1-z)$, and $\hat{\sigma}^{(1)incl}_{e^+e^- \rightarrow\gamma X}
(x_\gamma)\ddot{\otimes} D^{(0)}_{\gamma\rightarrow\gamma} (z)$
vanishes, the expression for the short-distance hard part becomes 
\begin{equation}
\hat{\sigma}^{(1)sub}_{e^+e^- \rightarrow \gamma X}(x_\gamma) =
  \left. \sigma^{(1)sub}_{e^+e^- \rightarrow \gamma X} 
  \right|_{E_q (or E_{\bar{q}})\geq \epsilon_h E_\gamma} 
- \hat{\sigma}^{(0)incl}_{e^+e^- \rightarrow q X} (x_q)
  \ddot{\otimes} D^{(1)}_{q\rightarrow \gamma} (z) 
- (q \rightarrow \bar{q})\ ,
\label{e16}
\end{equation}
where the terms with the minus sign are often called collinear
counter-terms. 
In this equation, $\hat{\sigma}^{(0)incl}_{e^+e^- \rightarrow q X}$ 
is obtained from Eq.~(\ref{e12}), and the modified convolution
``$\ddot{\otimes}$'' is defined in Eq.~(\ref{e8c}).  The perturbative
fragmentation function $D^{(1)}_{q\rightarrow \gamma} (z)$ in
$n$-dimensions can be calculated by evaluating the diagram in
Fig.~\ref{fig5}.  We obtain \cite{BGQ1} 
\begin{equation}
D^{(1)}_{q\rightarrow \gamma} (z) = 
e^2_q\ \left(\frac{\alpha_{em}}{2\pi}\right) 
\frac{1 + (1-z)^2}{z}\ \left( \frac{1}{-\epsilon} \right);
\label{e17}
\end{equation}
and $D^{(1)}_{\bar{q}\rightarrow \gamma} (z) = 
D^{(1)}_{q\rightarrow \gamma} (z)$.  Although
$\sigma^{(1)sub}_{e^+e^- \rightarrow\gamma X}$ and the perturbative 
fragmentation functions 
$D^{(1)}_{q ({\rm or}\, \bar{q})\rightarrow \gamma}(z)$
are both formally divergent as $\epsilon\rightarrow 0$, 
these divergences cancel
and leave a finite expression for 
$\hat{\sigma}^{(1)sub}_{e^+e^- \rightarrow \gamma X}$, 
if the conventional QCD factorization theorem holds \cite{AEMP}.

For the gluon fragmentation contribution: 
$e^+e^- \rightarrow g \rightarrow \gamma$ (Eq.~(\ref{e14c})), 
we apply Eq.~(\ref{e8}) to a gluon state perturbatively 
to first order in $\alpha_s$.  Letting $E_g$ be the gluon's energy, 
we obtain
\begin{eqnarray}
\left. \sigma^{(1)sub}_{e^+e^- \rightarrow gX} 
   \right|_{E_q (or E_{\bar{q}})\geq \epsilon_{\min}E_g }
&=&\hat{\sigma}^{(1)sub}_{e^+e^- \rightarrow gX} (x_g)
   \dot{\otimes} D^{(0)}_{g\rightarrow g} (z)
 + \hat{\sigma}^{(1)incl}_{e^+e^- \rightarrow gX} (x_g)
   \ddot{\otimes} D^{(0)}_{g\rightarrow g} (z)
\nonumber \\
&+&\hat{\sigma}^{(0)sub}_{e^+e^- \rightarrow qX} (x_q)
   \dot{\otimes} D^{(1)}_{q\rightarrow g} (z)
 + \hat{\sigma}^{(0)incl}_{e^+e^- \rightarrow q X} (x_q)
   \ddot{\otimes} D^{(1)}_{q\rightarrow g} (z)
\nonumber \\
&+&(q \rightarrow \bar{q})\ ,
\label{e18a}
\end{eqnarray}
where $x_g=2E_g/\sqrt{s}$, and $x_q=2E_q/\sqrt{s}$.  In
Eq.~(\ref{e18a}), the modified convolutions are defined in the same
way as in Eq.~(\ref{e8t}), except that $x_\gamma$ is replaced by
$x_g$, and $\epsilon_h$ is replaced by $\epsilon_{\min}$;
\begin{eqnarray}
\epsilon_{\min}(z) 
&\equiv & \frac{E_{\min}(z)}{E_g} \nonumber \\
&=& (1+\epsilon_h)z -1 \geq 0\ ,
\label{e18b}
\end{eqnarray}
where $z=x_\gamma/x_g$.  Because $z\leq 1$, we note the restrictions 
\begin{equation}
\epsilon_{\min}(z)\leq \epsilon_h\ ;
\quad \mbox{or} \quad
\frac{1}{1+\epsilon_{\min}} \geq \frac{1}{1+\epsilon_h}\ .
\label{e18c}
\end{equation}
Since $D^{(0)}_{g\rightarrow g} (z)=\delta(1-z)$, and 
$\hat{\sigma}^{(0)sub}_{e^+e^- \rightarrow qX} (x_q)=0$,  
Eq.~(\ref{e18a}) can be rewritten as
\begin{equation}
\hat{\sigma}^{(1)sub}_{e^+e^- \rightarrow g X}(x_g) = 
  \left. \sigma^{(1)sub}_{e^+e^- \rightarrow g X} 
  \right|_{E_q (or E_{\bar{q}})\geq \epsilon_{\min}E_g } 
- \hat{\sigma}^{(0)incl}_{e^+e^- \rightarrow qX} (x_q)
  \ddot{\otimes} D^{(1)}_{q \rightarrow g}
- (q \rightarrow \bar{q}).
\label{e18} 
\end{equation}
In Eq.~(\ref{e18}), the divergent cross section 
$\sigma^{(1)sub}_{e^+e^- \rightarrow g X}$
is evaluated from the Feynman graphs shown in Fig.~\ref{fig6}, 
and the quark-to-gluon
collinear divergences are embedded in the first-order 
fragmentation function $D^{(1)}_{q\rightarrow g}$.  
The function $D^{(1)}_{q\rightarrow g}$ is the same as 
$D^{(1)}_{q\rightarrow \gamma}$ of Eq.~(\ref{e17}), except that
$e_q^2(\alpha_{em}/2\pi)$ is replaced by $C_F(\alpha_s/2\pi)$.
The color factor $C_F=4/3$.

For the quark fragmentation contribution:
$e^+e^- \rightarrow q \rightarrow \gamma$ (Eq.~(\ref{e14b})),
we apply Eq.~(\ref{e8}) to a quark state perturbatively 
at first order in $\alpha_s$.  In a fashion similar to the derivation 
of $\hat{\sigma}^{(1)sub}_{e^+e^-\rightarrow gX}$, we 
derive the short-distance hard part for quark fragmentation
\begin{equation}
\hat{\sigma}^{(1)sub}_{e^+e^- \rightarrow q X}(x_q) = 
  \left. \sigma^{(1)sub}_{e^+e^- \rightarrow q X} 
  \right|_{E_g (or E_{\bar{q}})\geq \epsilon_{\min} E_q}
- \hat{\sigma}^{(0)incl}_{e^+e^- \rightarrow q'}(x_{q'}) 
  \ddot{\otimes} D^{(1)}_{q'\rightarrow q}(z) \ .
\label{e19}
\end{equation}
The Feynman graphs that contribute to 
$\sigma^{(1)sub}_{e^+e^- \rightarrow q X}$ 
are sketched in Fig.~\ref{fig7}. In this fragmentation process, 
the quark is effectively ``observed'' through $q\rightarrow \gamma$ 
fragmentation.  For the real gluon emission diagrams, sketched in 
Fig.~\ref{fig7}a, the gluon and antiquark are not observed and 
their momenta will be integrated over.  
Since $\epsilon_{\min}(z)$, defined in Eq.~(\ref{e18b}), 
can be equal to zero, the contribution of the real emission diagrams 
to $\sigma^{(1)sub}_{e^+e^- \rightarrow q X}$ shows both infrared
and collinear singularities.  The infrared singularity associated with 
soft gluon emission should be canceled by a contribution from  
the virtual gluon exchange diagrams shown in Fig.~\ref{fig7}b, if the
conventional QCD factorization theorem holds.
When the gluon is parallel to the fragmenting quark, the real emission 
diagrams manifest a collinear singularity that should be canceled 
by the negative term in Eq.~(\ref{e19}).

The perturbative fragmentation function
$D^{(1)}_{q\rightarrow q} (z)$ in Eq.~(\ref{e19}) is determined from 
Feynman diagrams sketched in Fig.~\ref{fig8}. In $n$-dimensions, we 
obtain \cite{BGQ1}
\begin{equation}
D^{(1)}_{q\rightarrow q} (z) = 
C_F \left({{\alpha_{s}} \over {2\pi}}\right) 
\left[ \frac{1 + z^2}{(1-z)_+}
+\frac{3}{2}\delta(1-z)\right] 
\left( {{1} \over {-\epsilon}}\right)\ ,
\label{e20}
\end{equation}
where the ``+'' description is defined as usual,
\begin{equation}
\left(\frac{1}{1-z}\right)_+ \equiv \frac{1}{1-z}
-\delta(1-z)\int_0^1 dz'\ \frac{1}{1-z'} .
\label{e20a}
\end{equation}

\subsection{Parton Level Cross Sections: $\sigma^{(1)sub}$ }
\label{subsec:3b}

In order to derive the short-distance hard parts at one-loop level, 
defined in Eqs.~(\ref{e16}), (\ref{e18}) and (\ref{e19}),
we must compute the formally divergent partonic
cross sections $\sigma^{(1)sub}_{e^+e^-\rightarrow cX}$ 
in $n$-dimensions, for $c=\gamma$, $g$, $q$ and $\bar{q}$. 
We must evaluate Feynman 
diagrams for $e^+ e^-$ to three particle final states, 
($q\bar{q}\gamma$) or ($q\bar{q}g$); and 
$e^+ e^-$ to two particle final states ($q\bar{q})$ with one-loop 
virtual gluon exchange.  As derived in Ref.~\cite{BGQ1}, 
the general expression for the partonic cross section is
\begin{equation}
d\sigma^{(1)sub} = \sum_{q,m} \left[\frac{2}{s}F^{PC}_q(s)\right]\ 
e^2\ C_q\ \frac{1}{4}(H_1+H_2)\ dPS^{(m)sub} ,
\label{e21}
\end{equation}
where $m=2$ or $3$ corresponding to the number of final state 
particles, and the constant $C_q$ is an overall color factor.  
In Eq.~(\ref{e21}), the functions $H_1$ and $H_2$ are defined by
\begin{equation}
H_1 = -g_{\mu\nu}\ H^{\mu\nu} , \  \  \  \
H_2 = - \frac{k_\mu k_\nu}{q^2}\ H^{\mu\nu} .
\label{e22}
\end{equation}
Function $H^{\mu\nu}$ is the hadronic tensor, and $q^\mu$ ($k^\mu$) is
the sum (difference) of the incoming $e^+$ and $e^-$ momenta. 
The factor $dPS^{(m)sub}$ in Eq.~(\ref{e21}) is the multi-particle
phase space element.  In $n=4-2\epsilon$ dimensions we obtain
\cite{BGQ1} 
\begin{equation}
dPS^{(2)sub} =
  \frac{1}{2}\frac{1}{(2\pi)^3}\frac{d^3p_i}{E_i}
  \left( \frac{4\pi}{(s/4) \sin^2\theta_i} \right)^\epsilon 
  \frac{1}{\Gamma(1-\epsilon)}
  \frac{2\pi}{s}
  \frac{\delta\left( x_i-1 \right)}{x_i} 
\label{e23a}
\end{equation}
for the two-body final state with parton $i$ being observed; and
\begin{eqnarray}
dPS^{(3)sub} &=&
  \frac{1}{2}\frac{1}{(2\pi)^3}\frac{d^3p_i}{E_i}
  \left( {{4\pi} \over {(s/4) \sin^2\theta_i}}\right)^\epsilon 
  {{1} \over {\Gamma(1-\epsilon)}}
  \frac{2\pi}{s}{{\delta\left( x_i-(1-y_{jh})\right)} \over {x_i}}
  \nonumber \\
&\times &\frac{s}{4}\left[
  \left(\frac{1}{2\pi}\right)^2\left(\frac{4\pi}{s}\right)^\epsilon
  \frac{1}{\Gamma(1-\epsilon)}
  \frac{d\Omega_{n-3}(p_j)}{\Omega_{n-3}} \right] \nonumber \\
&\times &{{dy_{12}}\over{y^\epsilon_{12}}}\, 
  {{dy_{13}}\over{y^\epsilon_{13}}}\, 
  {{dy_{23}}\over{y^\epsilon_{23}}}\, 
  \delta\left( 1-y_{12}-y_{13}-y_{23}\right),
\label{e23}
\end{eqnarray}
for the three-body final states.  In Eq.~(\ref{e23}), we use
labels ``1'', ``2'', and ``3'' for the final-state
$q$, $\bar{q}$ and $\gamma$ (or $g$), respectively; 
and we use label ``$i$'' (=1,2 or 3) to designate the ``observed''
one.  Symbol ``$j$''($\neq i$) labels the particle over whose momentum
we integrate, and the momentum of ``$h$''($\neq i\neq j$) is fixed by
the momentum conservation $\delta$-function.  In Eq.~(\ref{e23}), 
$\Omega_{n-3}=2\pi^{(1-\epsilon)}/\Gamma(1-\epsilon)$.
The dimensionless invariants $x_i$ and $y_{ij}$ in Eq.~(\ref{e23})
are defined as
\begin{equation}
x_i\equiv\frac{2E_i}{\sqrt{s}}\ , \quad\quad
y_{ij}\equiv \frac{2 p_i\cdot p_j}{s} ,
\label{e24}
\end{equation}
with $i,j = 1,2,3$.  

The multi-particle phase space expressions in Eqs.~(\ref{e23a}) 
and (\ref{e23}), are formally identical to those for our calculation
of the inclusive cross section \cite{BGQ1}, except that the limits of 
integration differ here, owing to the isolation condition.
These phase space limits are derived below for each individual
subprocess. 

As indicated in Figs.~\ref{fig4}, \ref{fig6} and \ref{fig7}, 
all three one-loop partonic cross sections have the same 
$e^+e^-$ to three-body matrix elements.  However, the quark 
fragmentation cross section has an extra contribution from virtual 
gluon exchange diagrams.  We derive the functions $H_1$ and $H_2$
appearing in Eq.~(\ref{e21}) by evaluating the square of the matrix
element for the Feynman diagrams in Fig.~\ref{fig4} (or
Fig.~\ref{fig6}, or Fig.~\ref{fig7}a).  
\begin{eqnarray}
H_1 &= & 8(1-\epsilon)\left\{(1-\epsilon)\left[
         \frac{y_{13}}{y_{23}}+\frac{y_{23}}{y_{13}}\right]
        +2\left[\frac{y_{12}}{y_{13}y_{23}} - \epsilon\right]\right\} ;
       \nonumber \\
H_2 &= & -4\left\{(1-\epsilon)\left[
         \frac{y_{13}}{y_{23}}+\frac{y_{23}}{y_{13}}\right]
        +2\left[\frac{y_{12}}{y_{13}y_{23}} - \epsilon\right]\right\} 
        \nonumber \\
    && + \frac{4}{y_{13}y_{23}}\left[ y^2_{1k} + y^2_{2k} \right]
       -\frac{4\epsilon}{y_{13}y_{23}}\ y^2_{3k} .
\label{e24a}
\end{eqnarray}
We temporarily omit overall coupling factors of 
$e_q^2\, (e\mu^\epsilon)^4$ 
for $(q\bar{q}\gamma)$ and $(e\mu^\epsilon)^2(g\mu^\epsilon)^2$ for 
$(q\bar{q}g)$.  The invariants $y_{ik}$ with $i=1,2,3$ are defined as
\begin{equation}
y_{ik}\equiv \frac{2p_i\cdot k}{s}.
\label{e24b}
\end{equation}
The identity 
\begin{equation}
y_{1k}+y_{2k}+y_{3k}=0
\label{yid}
\end{equation}
follows from the equalities $k\cdot q=0$ and $q=p_1+p_2+p_3$.

We choose to work in the overall center of mass frame, sketched in
Fig.~\ref{fig9}, when integrating $H_1$ and $H_2$ over three-body
phase space.  Letting $\theta_{ik}$ be the polar angle of the 
observed parton momentum $p_i$ with respect to the momentum $k$, 
and angle $\theta_{jx}$ the $n$-dimensional generalization of the 
three-dimensional azimuthal angle $\phi$, defined through $p_j$,
we derive general expressions for the three $y_{ik}$ \cite{BGQ1}.  
\begin{eqnarray}
y_{ik}&=& -x_i \cos\theta_{ik} ;\nonumber \\
y_{jk}&=& -\left[\frac{y_{ih}y_{jh}-y_{ij}}{x_i}\right]\cos\theta_{ik}
         -\left[\frac{2\sqrt{y_{ij}y_{ih}y_{jh}}}{x_i}\right]
            \sin\theta_{ik}\cos\theta_{jx} ;\nonumber \\
y_{hk}&=& -\left[\frac{y_{ij}y_{jh}-y_{ih}}{x_i}\right]\cos\theta_{ik}
         +\left[\frac{2\sqrt{y_{ij}y_{ih}y_{jh}}}{x_i}\right]
            \sin\theta_{ik}\cos\theta_{jx} .
\label{e24c}
\end{eqnarray}
These general expressions satisfy the identity in Eq.~(\ref{yid}).  
In the reference frame we are using,  
\begin{equation}
d\Omega_{n-3}(p_j) 
=d\cos\theta_{jx}\ (1-\cos^2\theta_{jx})^{n-5 \over 2}\ 
   d\Omega_{n-4}(p_j)\ .
\label{omega3}
\end{equation}

Since we choose the photon direction to define the $z$-axis (
Fig.~\ref{fig9}), with our cone definition, the range of the azimuthal
angle $\theta_{jx}$ integration should not be affected by the
isolation cut.  Therefore, just as the inclusive case \cite{BGQ1}, the
integral over $d\cos\theta_{jx}$ is 
done from $\cos\theta_{jx} = -1\ {\rm to}\ +1$ when $H_2$ is
integrated over phase space.  The expression for three-body phase
space, Eqs.~(\ref{e23}) and (\ref{omega3}), is an even
function of $\cos \theta_{jx}$.  
Correspondingly, terms in $H_2$ that are odd
functions of $\cos \theta_{jx}$  do not survive after integration.  
Because $H_2$ (Eq.~(\ref{e24a})) depends only on 
the square of $y_{1k}$ and $y_{2k}$, after eliminating all terms
linear in $\cos\theta_{jx}$, we find that 
the only remaining dependence on $\theta_{jx}$ in $H_2$ 
is of the form $\cos^2\theta_{jx}$.  We may integrate 
over $\theta_{jx}$ independently of other variables, or, equivalently, 
we may replace $\cos^2\theta_{jx}$ in $H_2$ by its average 
in $n$-dimensions and thereby eliminate $\theta_{jx}$ 
dependence in $H_2$ completely.  

The average of $\cos^2\theta_{jx}$ in $n$-dimensions is \cite{BGQ1}
\begin{equation}
<\cos^2\theta_{jx}>_{n-dim} = \frac{1}{2(1-\epsilon)} .
\label{e24d}
\end{equation}
We use Eq.~(\ref{e24d}) to replace $\cos^2\theta_{jx}$ in $H_2$ by
$1/2(1-\epsilon)$, and we denote the resulting expression $H_2^{eff}$.

In following subsections, for a given observed particle 
(e.g. $\gamma$ or $q$), we derive expressions for 
$y_{ik}^2$, for $i=1,2,3$,
substitute these $y_{ik}^2$ into Eq.~(\ref{e24a}), 
replace $\cos^2\theta_{jx}$ by $1/2(1-\epsilon)$
to obtain $(H_1+H_2^{eff})$, and finally, substitute 
these $(H_1+H_2^{eff})$ into Eq.~(\ref{e21}) to get the partonic 
cross sections: $\sigma^{(1)sub}$. 

\subsection{Derivation of 
$\hat{\sigma}^{(1)sub}_{e^+e^- \rightarrow \gamma X}$}
\label{subsec:3c}

In this subsection we present an explicit derivation of the finite 
hard-scattering cross section 
$\hat{\sigma}^{(1)sub}_{e^+e^- \rightarrow \gamma X}$ at
order $\alpha_{em}$, as expressed in Eq.~(\ref{e16}).

Since the photon is the observed particle, we set $i=3$.
The assignment of $j$ and $h$ is arbitrary and has no effect on the 
final result owing to symmetry between the quark and antiquark.
We choose $j=1$ and $h=2$; i.e., 
``1'' for the quark, and ``2'' for the antiquark.  Referring 
to Eq.~(\ref{e24}), we remark that $y_{13}$ is proportional to the 
invariant mass of the photon and quark system.  Starting from
Eq.~(\ref{e24c}), and after using Eq.~(\ref{e24d}), we obtain 
the effective expressions
\begin{eqnarray}
y_{3k}^2&=& \ x_3^2 \cos^2\theta_{3} ;\nonumber \\
y_{1k}^2&=& \left[\frac{y_{23}y_{12}-y_{13}}{x_3}\right]^2
           \cos^2\theta_{3}
         +\frac{1}{1-\epsilon}
            \left[\frac{2(y_{12}y_{13}y_{23})}{x_3^2}\right]
            \sin^2\theta_{3};\nonumber \\
y_{2k}^2&=& \left[\frac{y_{13}y_{12}-y_{23}}{x_3}\right]^2
           \cos^2\theta_{3}
         +\frac{1}{1-\epsilon}
            \left[\frac{2(y_{12}y_{13}y_{23})}{x_3^2}\right]
            \sin^2\theta_{3} .
\label{e24e}
\end{eqnarray}
Substituting these $y_{ik}^2$, with $i=1,2,3$, into Eq.~(\ref{e24a}), 
we derive
\begin{eqnarray}
{{1} \over {4}} \left( H_1+H_2^{eff} \right) 
= e_q^2 (e\mu^\epsilon)^4 
&\Bigg\{ &
  \left( 1 + \cos^2 \theta_3 - 2\epsilon\right)
  \left[ (1-\epsilon)
        \left(\frac{y_{13}}{y_{23}}+\frac{y_{23}}{y_{13}}\right)
       + 2 \left(\frac{y_{12}}{y_{13}y_{23}}-\epsilon\right) \right] 
\nonumber \\
&+& \left( 1-3\cos^2\theta_3\right) 
   \left[\frac{4y_{12}}{x_3^2}\right] \Bigg\} \nonumber \\
= e_q^2 (e\mu^\epsilon)^4 
&\Bigg\{ &
  \left( 1 + \cos^2 \theta_3 - 2\epsilon\right)
  \left[ {{1+(1-x_3)^2} \over {x^2_3}}\right]
  \left( {{1} \over {\hat{y}_{13}}} + {{1} \over
         {\hat{y}_{23}}}\right)
\nonumber \\
&+& \left( 1 + \cos^2\theta_3 - 2\epsilon\right) 
  \left[ -2-\epsilon\left( {{1}\over{\hat{y}_{13}}}+
  {{1}\over {\hat{y}_{23}}}\right)\right]\nonumber \\
&+& \left( 1-3\cos^2\theta_3\right) 
  \left[ {{4(1-x_3)} \over {x^2_3}} \right] \Bigg\} .
\label{e25}
\end{eqnarray}
In Eq.~(\ref{e25}), we introduce the overall coupling factor and 
neglect terms that do not contribute in the limit 
$\epsilon\rightarrow 0$. 
The variable $\hat{y}_{j3}$ with $j=1,2$
is defined as $\hat{y}_{j3} = y_{j3}/x_3$; $x_3$
is the same as $x_\gamma$, and $\theta_3$ is equal to $\theta_\gamma$
in this case.

The two $\delta$ functions in $dPS^{(3)}$, Eq.~(\ref{e23}),
are used to do the $dy_{12}$ and $dy_{23}$ integrations, 
and $y_{13}$ is left as the integration variable.  Different 
choices of the integration variable are of course equivalent.  In 
terms of $y_{13}$ and $x_3$, we have identities
\begin{equation}
y_{12} = 1 - x_3 , \quad \quad
y_{23} = x_3 - y_{13} .
\label{e26}
\end{equation}

The limits of integration over $y_{13}$ are derived from the isolation
requirement for the subtraction terms.  In our parton level approach, 
hadronic energy in the isolation cone means a parton must be present 
inside the isolation cone of observed photon.  Second, this hadronic
energy must be larger than $\epsilon_h E_\gamma$.
We use $\delta$ to denote the half angle of the isolation cone.  The
statement that the quark (parton ``1'') is inside the isolation cone
provides the relationship
\begin{equation}
\hat{y}_{23} \geq \frac{\cos^2(\delta/2)}{1-x_3 \sin^2(\delta/2)} 
\label{e27} ,
\end{equation}
whereas the statement that the antiquark (parton ``2'') is inside the 
isolation cone leads to
\begin{equation}
\hat{y}_{13} \geq \frac{\cos^2(\delta/2)}{1-x_3 \sin^2(\delta/2)} .
\label{e28} 
\end{equation}
Using Eq.~(\ref{e26}) and $0\leq y_{13} \leq x_3$, 
we derive the condition that either the quark or the  
antiquark is inside the isolation cone:
\begin{eqnarray} 
&&0 \quad \leq \quad
   \hat{y}_{13} \quad \leq \quad 
\frac{(1-x_3)\sin^2(\delta/2)}{1-x_3 \sin^2(\delta/2)}\ ; \nonumber \\
&& \frac{\cos^2(\delta/2)}{1-x_3 \sin^2(\delta/2)} 
\quad \leq \quad 
 \hat{y}_{13} \quad \leq \quad 1 .
\label{e29}
\end{eqnarray}
These two regions of the $\hat{y}_{13}$ integration do not
overlap as long as $x_3>1-\cot^2(\delta/2)$.

The second part of the isolation constraint is the requirement that 
the parton's energy inside the isolation cone be larger than 
$\epsilon_h E_\gamma$.  We derive 
\begin{eqnarray} 
{\rm max}[0,(1+\epsilon_h -1/x_3)] \quad \leq \quad 
& \hat{y}_{13}& \quad \leq \quad  
\frac{(1-x_3)\sin^2(\delta/2)}{1-x_3 \sin^2(\delta/2)}\ , 
\nonumber \\
1-\frac{(1-x_3)\sin^2(\delta/2)}{1-x_3 \sin^2(\delta/2)} 
\quad \leq \quad 
& \hat{y}_{13}& \quad \leq \quad {\rm min}[1,1-(1+\epsilon_h-1/x_3)]\ .
\label{e30}
\end{eqnarray}
For simplicity of notation and to facilitate comparison of our results
with those of Ref.~\cite{KT}, we define
\begin{eqnarray}
y_c &\equiv &\frac{(1-x_3)\sin^2(\delta/2)}{1-x_3 \sin^2(\delta/2)} 
     \quad \Rightarrow \quad 
     (1-x_3)\frac{\delta^2}{4}, \nonumber \\
y_m &\equiv &1+\epsilon_h - \frac{1}{x_3} .
\label{e31}
\end{eqnarray}
Henceforth in this subsection, we set $x_3=x_\gamma$.  The first 
of the conditions in Eq.~(\ref{e30}) indicates that 
$y_m$ should be less than $y_c$.  Therefore, the subtraction
term should vanish if $y_m\geq y_c$, i.e., 
\begin{equation}
E_\gamma\frac{\hat{\sigma}^{(1)sub}_{e^+e^-\rightarrow\gamma X}}
             {d^3\ell} = 0
\quad
\mbox{if} \quad 
x_\gamma\geq x_\gamma^{\rm max}(\delta,\epsilon_h)\ .
\label{e32}
\end{equation}
Here, $x_\gamma^{\rm max}(\delta,\epsilon_h)$ is expressed as 
\begin{equation}
x_\gamma^{\rm max}(\delta,\epsilon_h) \equiv
\left(\frac{1}{1+\epsilon_h}\right)
  \frac{2}{1+\sqrt{1-4\epsilon_h\sin^2(\delta/2)/(1+\epsilon_h)^2}} .
\label{e32a}
\end{equation}
We note that 
\begin{equation}
x_\gamma^{\rm max}(\delta,\epsilon_h) 
\geq \frac{1}{1+\epsilon_h}
\label{e32c}
\end{equation}
if $\delta\neq 0$.  We define 
\begin{equation}
\Delta 
\equiv x_{\gamma}^{\rm max}(\delta,\epsilon_h) 
     - \frac{1}{1+\epsilon_h}
\simeq \epsilon_h\,\frac{\delta^2}{4} \ ,
\label{eDelta}
\end{equation} 
where we expand $x_\gamma^{\rm max}(\delta,\epsilon_h)$ of
Eq.~(\ref{e32c}) for small $\delta$.  For $\delta=20^\circ$, typical
of LEP experiments, $\Delta\simeq 0.03\epsilon_h$.  
The interval between the critical value $x_\gamma =
1/(1+\epsilon_h)$ and $x_\gamma^{\rm max}$ is consequently very small.
Equation~(\ref{e32}) is actually a condition due to energy-momentum
conservation. 

We consider next, in turn, the regions $x_\gamma \leq
1/(1+\epsilon_h)$ and $x_\gamma > 1/(1+\epsilon_h)$.  
If $x_\gamma \leq 1/(1+\epsilon_h)$, the integration over
$\hat{y}_{13}$ for $d\sigma^{(1)sub}$ has two separate intervals,
defined in Eq.~(\ref{e30}), 
\begin{equation}
\int d\hat{y}_{13} 
\Rightarrow \int_{0}^{y_c} d\hat{y}_{13} + 
  \int_{1-y_c}^{1} d\hat{y}_{13} .
\label{e32b}
\end{equation}
Substituting Eq.~(\ref{e25}) into Eq.~(\ref{e21}), and performing 
this $\hat{y}_{13}$ integration, we derive parton level cross section 
\begin{eqnarray}
&E_\gamma &
\frac{d{\sigma}^{(1)sub}_{e^+e^-\rightarrow\gamma X}}{d^3\ell}
= \, 2 \sum_{q} \left[\frac{2}{s}F^{PC}_{q}(s)\right]
 \left[\alpha_{em}^2N_c
 \left(\frac{4\pi\mu^2}{(s/4)\sin^2\theta_\gamma}\right)^\epsilon
       \frac{1}{\Gamma(1-\epsilon)}\right] \frac{1}{x_\gamma}\,
       e_q^2 \left(\frac{\alpha_{em}}{2\pi}\right) 
\nonumber \\
&\quad &
\times \left\{ (1+\cos^2\theta_\gamma-2\epsilon)
                 \left(\frac{1+(1-x_\gamma)^2}{x_\gamma}\right)
                 \left(-\frac{1}{\epsilon}\right) \right. 
\nonumber \\
&\quad & \quad
+ (1+\cos^2\theta_\gamma)
  \left[\left(\frac{1+(1-x_\gamma)^2}{x_\gamma}\right)
        \left(\ell n\left(\frac{s}{\mu^2_{\overline{\rm MS}}}\right)
            + \ell n(x_\gamma^2(1-x_\gamma)) 
            + \ell n(y_c) + y_c   \right) \right. 
\nonumber \\
&\quad & \quad
\mbox{\hskip 1.2in} 
+ x_\gamma(1-2y_c) \Bigg]  
\nonumber \\
&\quad & \quad
+ \left. (1-3\cos^2\theta_\gamma)
         \left[2\left(\frac{1-x_\gamma}{x_\gamma}\right)\right]
         \left(2y_c\right) \right\} .
\label{e35}
\end{eqnarray}
In Eq.~(\ref{e35}), the usual modified minimal subtraction scale is
\begin{equation}
\mu_{\overline{\rm MS}}^2 =\mu^2\ 4\pi\ e^{-\gamma_E} ,
\label{mumsb}
\end{equation} 
where $\gamma_E$ is Euler's constant.  The $1/\epsilon$ poles in
Eq.~(\ref{e35}) arise from the $1/\hat{y}_{13}$ 
and $1/\hat{y}_{23}$ terms in Eq.~(\ref{e25}).  Referring to
Eq.(\ref{e16}), and using Eqs.~(\ref{e12}) and (\ref{e17}), we observe
that the $1/\epsilon$ pole in Eq.~(\ref{e35}) is exactly canceled. 
Consequently, for $x_\gamma \leq 1/(1+\epsilon_h)$, the finite
short-distance hard part is
\begin{eqnarray}
E_\gamma
\frac{d\hat{\sigma}^{(1)sub}_{e^+e^-\rightarrow\gamma X}}{d^3\ell}
&= & \, 2 \sum_{q} \left[\frac{2}{s}F^{PC}_{q}(s)\right]
               \left[\alpha_{em}^2N_c \frac{1}{x_\gamma} \right]
          e_q^2 \left(\frac{\alpha_{em}}{2\pi}\right) \nonumber \\
&\times & (1+\cos^2\theta_\gamma)
    \left\{\left(\frac{1+(1-x_\gamma)^2}{x_\gamma}\right)
    \left[\ell n\left(\frac{s}{\mu^2_{\overline{\rm MS}}}\right)
        + \ell n(x_\gamma^2(1-x_\gamma)) \right.\right.
               \nonumber \\
&& \mbox{\hskip 1.5in} \left. \left. 
                + \ell n\left((1-x_\gamma)
                        \frac{\delta^2}{4}\right) \right]
                + x_\gamma \right\} .
\label{e36}
\end{eqnarray}
In Eq.~(\ref{e36}), we neglect terms of $O(\delta^2)$.  

If $x_\gamma > 1/(1+\epsilon_h)$, the negative term, defined through 
the $\ddot{\otimes}$ convolution in Eq.~(\ref{e16}), vanishes. In
this region of phase space, there should not be any collinear
divergences in the partonic cross section, 
$E_\gamma d{\sigma}^{(1)sub}_{e^+e^-\rightarrow\gamma X}/d^3\ell$,
since there is no counter term to cancel them.
If the fragmentation process were exactly collinear, the subtraction
term $E_\gamma d{\sigma}^{(1)sub}_{e^+e^-\rightarrow\gamma X}/d^3\ell$ 
would vanish, kinematically, for $x_\gamma > 1/(1+\epsilon_h)$.  
However, a finite isolation cone $\delta\neq 0$ allows for a
non-vanishing $E_\gamma d{\sigma}^{(1)sub}_{e^+e^-\rightarrow\gamma
X}/d^3\ell$ even when $x_\gamma > 1/(1+\epsilon_h)$.
Equation~(\ref{e32}) shows that there is a narrow interval, 
$1/(1+\epsilon_h)<x_\gamma<x_\gamma^{\rm max}(\delta,\epsilon_h)$,
in which a non-vanishing 
$E_\gamma d{\sigma}^{(1)sub}_{e^+e^-\rightarrow\gamma X}/d^3\ell$
is allowed kinematically.

In the narrow interval, 
$1/(1+\epsilon_h)<x_\gamma<x_\gamma^{\rm max}(\delta,\epsilon_h)$,
the integration over $\hat{y}_{13}$ for $d\sigma^{(1)sub}$ has two 
separate regions:
\begin{equation}
\int d\hat{y}_{13} 
\Rightarrow \int_{y_m}^{y_c} d\hat{y}_{13} + 
  \int_{1-y_c}^{1-y_m} d\hat{y}_{13}\ .
\label{e33}
\end{equation}
Equation~(\ref{e31}) shows that $y_m>0$.  Therefore, 
$\hat{y}_{13} > 0$ and $\hat{y}_{13}<1$.  Consequently, there is no
collinear divergence and the integration
over $\hat{y}_{13}$ can be done completely in $n=4$ dimensions.  

For $1/(1+\epsilon_h)<x_\gamma<x_\gamma^{\rm max}(\delta,\epsilon_h)$,
we derive
\begin{eqnarray}
E_\gamma\frac{d\hat{\sigma}^{(1)sub}_{e^+e^-\rightarrow\gamma X}}
             {d^3\ell}
&= & \, 2 \sum_{q} \left[\frac{2}{s}F^{PC}_{q}(s)\right]
               \left[\alpha_{em}^2N_c\ \frac{1}{x_\gamma}\right]
               e_q^2 \left(\frac{\alpha_{em}}{2\pi}\right) 
\nonumber \\
&& \times \left\{ (1+\cos^2\theta_\gamma)
         \left[\left(\frac{1+(1-x_\gamma)^2}{x_\gamma}\right)
               \left(\ell n \frac{y_c}{y_m} +
                     \ell n \frac{1-y_m}{1-y_c}\right) \right. \right. 
               \nonumber \\
&& \mbox{\hskip 1.2in} %\left.
               -4x_\gamma\left(y_c - y_m\right) \Bigg] 
\nonumber \\
&& \quad + \left. (1-3\cos^2\theta_\gamma)
         \left[4\left(\frac{1-x_\gamma}{x_\gamma}\right)\right]
         \left(y_c-y_m\right) \right\} 
\nonumber \\
&= & \, 2 \sum_{q} \left[\frac{2}{s}F^{PC}_{q}(s)\right]
               \left[\alpha_{em}^2N_c\ \frac{1}{x_\gamma} \right]
               e_q^2 \left(\frac{\alpha_{em}}{2\pi}\right)  
\nonumber \\
&& \times (1+\cos^2\theta_\gamma)
         \left[\left(\frac{1+(1-x_\gamma)^2}{x_\gamma}\right)
               \ell n \left( 
                \frac{(1-x_\gamma)\delta^2/4}{1+\epsilon_h-1/x_\gamma} 
               \right) \right] 
\nonumber \\
&& + O(\delta^2) .
\label{e34}
\end{eqnarray}

We remark that 
$E_\gamma d\hat{\sigma}^{(1)sub}_{e^+e^-\rightarrow\gamma X}/d^3\ell$
in Eq.~(\ref{e34}) manifests a logarithmic divergence, of the nature of
a collinear divergence, as $x_\gamma\rightarrow 1/(1+\epsilon_h)$.
This problem arises from the incompatibility of collinear fragmentation, 
used to define the counter term in Eq.~(\ref{e16}),
and the cone fragmentation used in the definition of the partonic
cross section $E_\gamma
d\hat{\sigma}^{(1)sub}_{e^+e^-\rightarrow\gamma X}/d^3\ell$ 
in Eq.~(\ref{e16}).  More discussion of this issue is presented below, 
in Sec.~\ref{subsec:4c}.  

\subsection{Derivation of 
$\hat{\sigma}^{(1)sub}_{e^+e^- \rightarrow g X}$}
\label{subsec:3d}

In this section we present our explicit expression for the finite 
short-distance hard part for the order $\alpha_{s}$ gluon
fragmentation process, $e^+e^- \rightarrow g \rightarrow \gamma$,
expressed in Eq.(\ref{e18}).  The similarity between the Feynman
diagrams for $e^+e^- \rightarrow \gamma$, shown in Fig.~\ref{fig4},
and those for $e^+e^- \rightarrow g \rightarrow \gamma$, shown in
Fig.~\ref{fig6} allows us to exploit the results derived in the
previous subsection. 

We derive $\hat{\sigma}^{(1)sub}_{e^+e^- \rightarrow gX}$ from the
expression for $\hat{\sigma}^{(1)sub}_{e^+e^- \rightarrow \gamma X}$,
given in Eqs.~(\ref{e36}) and (\ref{e34}), by making the 
following four replacements: $x_\gamma \rightarrow x_g$;
$N_c\rightarrow N_c C_F$; 
$e^2 e_q^2$ of the final photon emission vertex by $g^2=4\pi\alpha_s$; 
and $\epsilon_h\rightarrow \epsilon_{\min}(z)$.  The last replacement 
is absent for the inclusive case.  Here $z\equiv x_\gamma/x_g$.  

If $x_g\leq 1/(1+\epsilon_{\min})$, which is the same as  
$x_\gamma \leq 1/(1+\epsilon_h)$, these replacements in
Eq.~(\ref{e36}) provide the finite short-distance hard part
\begin{eqnarray}
E_g
\frac{d\hat{\sigma}^{(1)sub}_{e^+e^-\rightarrow gX}}{d^3p_g}
&= & \, 2 \sum_{q} \left[\frac{2}{s}F^{PC}_{q}(s)\right]
               \left[\alpha_{em}^2N_c \frac{1}{x_g} \right]
               C_F \left(\frac{\alpha_s}{2\pi}\right) \nonumber \\
&& \times (1+\cos^2\theta_3)
          \left\{\left(\frac{1+(1-x_g)^2}{x_g}\right)
          \left[\ell n\left(\frac{s}{\mu^2_{\overline{\rm MS}}}\right)
                    + \ell n(x_g^2(1-x_g)) \right.\right.\nonumber \\
&& \mbox{\hskip 1.5in} \left.\left.
                         + \ell n\left((1-x_g)
                                 \frac{\delta^2}{4}\right) \right]
                + x_g \right\} \nonumber \\
&& + O(\delta^2) \ .
\label{e37}
\end{eqnarray}

If $x_g > 1/(1+\epsilon_{\min})$, which is the same as  
$x_\gamma > 1/(1+\epsilon_h)$, we make the four replacements in
Eq.~(\ref{e34}) and find
\begin{eqnarray}
E_g
\frac{d\hat{\sigma}^{(1)sub}_{e^+e^-\rightarrow gX}}{d^3p_g}
&= &\, 2 \sum_{q} \left[\frac{2}{s}F^{PC}_{q}(s)\right]
               \left[\alpha_{em}^2N_c\ \frac{1}{x_g} \right]
               C_F \left(\frac{\alpha_s}{2\pi}\right)  \nonumber \\
&& \times (1+\cos^2\theta_3)
         \left[\left(\frac{1+(1-x_g)^2}{x_g}\right)
               \ell n \left( 
                 \frac{(1-x_g)\delta^2/4}
                      {(1+\epsilon_h)(x_\gamma/x_g)-1/x_g} 
               \right) \right] \nonumber \\
&& + O(\delta^2) .
\label{e38}
\end{eqnarray}

From energy-momentum conservation, we derive an equation similar to 
Eq.~(\ref{e32}), 
\begin{equation}
E_g\frac{d\hat{\sigma}^{(1)sub}_{e^+e^-\rightarrow gX}}{d^3p_g}
= 0
\quad \mbox{if} \quad
x_g \geq x^{\rm max}(z,\delta,\epsilon_h) .
\label{e39}
\end{equation}
The maximum value $x^{\max}$ is  
\begin{equation}
x^{\rm max}(z,\delta,\epsilon_h) \equiv
  \left(\frac{1}{z(1+\epsilon_h)}\right)
  \frac{2}{1+\sqrt{1-\frac{4\sin^2(\delta/2)(z(1+\epsilon_h)-1)}
                          {z^2(1+\epsilon_h)^2}}} \ .
\label{e44a}
\end{equation}

The gluon fragmentation contribution to the $O(\alpha_s)$ subtraction
term in $\sigma^{(1)sub}_{e^+e^-\rightarrow\gamma X}$ is provided by
the convolution 
\begin{equation}
E_\gamma 
\frac{d\sigma^{(1)sub}_{e^+e^-\rightarrow gX\rightarrow \gamma X}}
     {d^3\ell}
 = \int^1_{\max\left[x_\gamma,\frac{1}{1+\epsilon_h}\right]}\, 
   \frac{dz}{z}
\left[ E_g \frac{d\hat{\sigma}_{e^+e^-\rightarrow gX}^{(1)sub}}
                {d^3p_g}
           \left(x_g=\frac{x_\gamma}{z}\right) \right]
\frac{D_{g\rightarrow\gamma}(z, \mu^2_F)}{z}\ .
\label{e44b}
\end{equation}
If collinear fragmentation is assumed for $g\rightarrow \gamma$,
$\theta_3=\theta_\gamma$.  We remark that, after convolution with the
gluon-to-photon fragmentation function, the subtraction term 
$E_g d\hat{\sigma}^{(1)sub}_{e^+e^-\rightarrow gX}/d^3p_g$ in 
Eq.~(\ref{e38}) develops a logarithmic divergence $\ell 
n(1+\epsilon_h-1/x_\gamma)$, as $x_\gamma \rightarrow 1/(1+\epsilon_h)$.
More discussion of this issue is presented below.

\subsection{Derivation of 
\protect $\hat{\sigma}^{(1)sub}_{e^+e^- \rightarrow q X}$ }
\label{subsec:3e}

In this section we evaluate the finite short-distance hard part
$\hat{\sigma}^{(1)sub}_{e^+e^- \rightarrow qX}$ at order 
$\alpha_{s}$, defined in Eq.~(\ref{e19}), for the quark fragmentation
process $e^+e^- \rightarrow q \rightarrow \gamma$.
Because $z\geq\max[x_\gamma,1/(1+\epsilon_h)]$ in the convolutions of 
Eq.~(\ref{e9a}), 
our general analysis of Sec.~\ref{sec:2} shows that 
the short-distance hard part $\hat{\sigma}^{(1)sub}_{e^+e^-
\rightarrow qX}$ is needed only for $x_1 = x_c \leq
\min[x_\gamma(1+\epsilon_h), 1]$.  We recall that $x_1=2E_q/\sqrt{s}$.
We analyze, in turn, the intervals 
$x_\gamma < 1/(1+\epsilon_h)$ and $x_\gamma > 1/(1+\epsilon_h)$.
Discussion of the special case $x_\gamma=1/(1+\epsilon_h)$ is reserved 
for Section~\ref{subsec:4d}. 

If $x_\gamma < 1/(1+\epsilon_h)$, then $x_1 < 1$ for all values of
$z$ in the convolution.  In this interval, the virtual diagrams of 
Fig.~\ref{fig7}b, whose contribution is proportional to
$\delta(1-x_1)$, do not contribute to $\hat{\sigma}^{(1)sub}_{e^+e^-
\rightarrow qX}$.   
The square of the matrix element for 
$\hat{\sigma}^{(1)sub}_{e^+e^- \rightarrow qX}$ 
is therefore the same as that in Eq.~(\ref{e24a}), 
except that the $y^2_{ik}$ variables are no longer those of 
Eq.~(\ref{e24e}).  Because the quark is now the fragmenting parton,
the labels are instead $i=1$, $j=3$, and $h=2$.  We obtain
\newpage
\begin{eqnarray}
y_{1k}^2&=& \ x_1^2 \cos^2\theta_{1} ;\nonumber \\
y_{2k}^2&=& \left[\frac{y_{13}y_{23}-y_{12}}{x_1}\right]^2
           \cos^2\theta_{1}
         +\frac{1}{1-\epsilon}
            \left[\frac{2(y_{12}y_{13}y_{23})}{x_1^2}\right]
            \sin^2\theta_{1}\ ;\nonumber \\
y_{3k}^2&=& \left[\frac{y_{12}y_{23}-y_{13}}{x_1}\right]^2
           \cos^2\theta_{1}
         +\frac{1}{1-\epsilon}
            \left[\frac{2(y_{12}y_{13}y_{23})}{x_1^2}\right]
            \sin^2\theta_{1}\ .
\label{e45}
\end{eqnarray}
In Eq.~(\ref{e45}), $\theta_1$ is the angle between the 
quark momentum $p_1$ and the momentum $k$.
In deriving Eq.~(\ref{e45}), we dropped terms linear 
in $\cos\theta_{jx}$, and replaced  
$\cos^2\theta_{jx}$ by its average in $n$-dimensions,
$1/2(1-\epsilon)$.
Substituting these $y^2_{ik}$ into Eq.~(\ref{e24a}), and using the
identities $y_{23}=1-x_1$ and $y_{12}=x_1-y_{13}$, we derive
\begin{eqnarray}
\frac{1}{4} \left( H_1+H_2^{eff} \right) 
= (e\mu^\epsilon)^2 (g\mu^\epsilon)^2 
&\Bigg\{ &
  \left( 1 + \cos^2 \theta_1 - 2\epsilon\right)
  \left[ (1-\epsilon)
        \left(\frac{y_{13}}{y_{23}}+\frac{y_{23}}{y_{13}}\right)
       + 2 \left(\frac{y_{12}}{y_{13}y_{23}}-\epsilon\right) \right] 
        \nonumber \\
&+& \left( 1-3\cos^2\theta_1\right) 
   \left[\frac{2y_{12}}{x_1^2}\right] \Bigg\} \nonumber \\
= (e\mu^\epsilon)^2 (g\mu^\epsilon)^2 
&\Bigg\{ &
  \left( 1 + \cos^2 \theta_1 - 2\epsilon\right)
  \left[ \left( \frac{1+x_1^2}{1-x_1} \right) \frac{1}{y_{13}} 
  + \frac{y_{13}}{1-x_1} \right] \nonumber \\
&+& \left( 1 + \cos^2\theta_1 - 2\epsilon\right) 
  \left[ -\frac{2}{1-x_1}
         -\epsilon\left( \frac{1-x_1}{y_{13}}+
                         \frac{y_{13}}{1-x_1}+2 \right) \right]
         \nonumber \\
&+& \left( 1-3\cos^2\theta_1\right) 
  \left[ \frac{2}{x_1} \left(1-\frac{y_{13}}{x_1}\right) \right] 
  \Bigg\}\ .
\label{e46}
\end{eqnarray}
The overall coupling constant $(e\mu^\epsilon)^2 (g\mu^\epsilon)^2$ is
restored in Eq.~(\ref{e46}).  Since $x_1 < 1$ for all values of $z$,
there is no infrared divergence associated with $1/(1-x_1)$ terms in
Eq.~(\ref{e46}). 

The isolation condition for the partonic cross section 
$\sigma^{(1)sub}_{e^+e^-\rightarrow qX}$ requires that
either the gluon or the antiquark be in the isolation cone of
the quark and have energy larger than $E_{\min}$ 
defined in Eq.~(\ref{e7}).  Following an analysis similar to the
one that led to Eq.~(\ref{e30}), 
we derive the limits of the integration over $\hat{y}_{13}\equiv
y_{13}/x_1$: 
\begin{eqnarray} 
{\rm max}[0,\bar{y}_m] \quad \leq \quad 
& \hat{y}_{13} & \quad \leq \quad  \bar{y}_c; \nonumber \\
1-\bar{y}_c \quad \leq \quad 
& \hat{y}_{13} & \quad \leq \quad {\rm min}[1,1-\bar{y}_m] \ .
\label{e47}
\end{eqnarray}
In Eq.~(\ref{e47}), $\bar{y}_c$ and $\bar{y}_m$ are defined as  
\begin{eqnarray}
\bar{y}_c &\equiv & 
     \frac{(1-x_1)\sin^2(\delta/2)}{1-x_1 \sin^2(\delta/2)} 
     \quad \Rightarrow \quad (1-x_1)\frac{\delta^2}{4}; \nonumber \\
\bar{y}_m &\equiv &
     (1+\epsilon_h)z - \frac{1}{x_1} ;
\label{e48}
\end{eqnarray}
with $z=x_\gamma/x_1$.  In analogy to Eq.~(\ref{e39}), 
\begin{equation}
E_1\frac{d\hat{\sigma}^{(1)sub}_{e^+e^-\rightarrow qX}}{d^3p_1}
= 0
\quad
\mbox{if} \quad 
x_1\geq x_{\rm max}(z,\delta,\epsilon_h)\ ,
\label{e49}
\end{equation}
where $x_{\rm max}(z,\delta,\epsilon_h)$ is defined in Eq.~(\ref{e44a}).  

In the region $x_\gamma < 1/(1+\epsilon_h)$, $\bar{y}_m <0$.  The
integration over $\hat{y}_{13}$ has two separate intervals: 
\begin{equation}
\int d\hat{y}_{13} 
\Rightarrow \int_{0}^{\bar{y}_c} d\hat{y}_{13} + 
  \int_{1-\bar{y}_c}^{1} d\hat{y}_{13}\ .
\label{e50}
\end{equation}
Integrating over $\hat{y}_{13}$, we obtain
\begin{eqnarray}
E_1\frac{d{\sigma}^{(1)sub}_{e^+e^-\rightarrow q X}}{d^3p_1}
&= & \, \left[\frac{2}{s}F^{PC}_{q}(s)\right]
    \left[\alpha_{em}^2N_c
          \left(\frac{4\pi\mu^2}{(s/4)\sin^2\theta}\right)^\epsilon
          \frac{1}{\Gamma(1-\epsilon)}\right] \frac{1}{x_1}
          C_F\left(\frac{\alpha_{s}}{2\pi}\right) \nonumber \\
&\times & \Bigg\{ (1+\cos^2\theta_1-2\epsilon)
                 \left(\frac{1+x_1^2}{1-x_1}\right)
                 \left(-\frac{1}{\epsilon}\right) \nonumber \\
&& + (1+\cos^2\theta_1)
               \Bigg[\left(\frac{1+x_1^2}{1-x_1}\right)
               \left(\ell n\left(\frac{s}{\mu^2_{\overline{\rm MS}}}\right)
                         + \ell n(x_1^2(1-x_1))
                         + \ell n\left((1-x_1)\frac{\delta^2}{4}\right)
                           \right)  \nonumber \\
       && \mbox{\hskip 1.2in}
              + (1-x_1) \Bigg]  \Bigg\}\ .
\label{e51}
\end{eqnarray}
Terms of $O(\delta^2)$ are neglected.  The $1/\epsilon$ pole 
in Eq.~(\ref{e51}) arises from
the $1/y_{13}$ term in Eq.~(\ref{e46}), corresponding to 
the collinear singularity when the gluon is 
parallel to the fragmenting quark.  Combining Eqs.~(\ref{e12}) and
Eq.~(\ref{e20}), and using the fact that $x_1<1$, one may verify 
that this collinear singularity is canceled exactly by the 
subtraction term in Eq.~(\ref{e19}).  

For the finite short-distance subtraction term in the region
$x_\gamma < 1/(1+\epsilon_h)$, we obtain 
\begin{eqnarray}
E_1\frac{d\hat{\sigma}^{(1)sub}_{e^+e^-\rightarrow q X}}{d^3p_1}
&= & \, \left[\frac{2}{s}F^{PC}_{q}(s)\right]
               \left[\alpha_{em}^2N_c\ \frac{1}{x_1}\right]
               C_F \left(\frac{\alpha_{s}}{2\pi}\right) \nonumber \\
&\times & (1+\cos^2\theta_1)
               \left\{\left(\frac{1+x_1^2}{1-x_1}\right)
               \left[\ell n\left(\frac{s}{\mu^2_{\overline{\rm MS}}}\right)
                         + \ell n(x_1^2(1-x_1)) \right.\right.\nonumber \\
&& \mbox{\hskip 1.5in} \left. \left. 
                         + \ell n\left((1-x_1)
                                  \frac{\delta^2}{4}\right) \right]
                + (1-x_1) \right\} .
\label{e52}
\end{eqnarray}

Turning to the case $x_1 > 1/(1+\epsilon_{\min(z)})$, which is the
same as $x_\gamma > 1/(1+\epsilon_h)$, we note that $\bar{y}_m > 0$.
The integration region over $\hat{y}_{13}\equiv y_{13}/x_1$ now has
the form  
\begin{equation}
\int d\hat{y}_{13} 
\Rightarrow \int_{\bar{y}_m}^{\bar{y}_c} d\hat{y}_{13} + 
  \int_{1-\bar{y}_c}^{1-\bar{y}_m} d\hat{y}_{13} .
\label{e53}
\end{equation}

Equation~(\ref{e49}) shows that $x_1 < 1$ for a nonvanishing
$E_1 d\hat{\sigma}^{(1)sub}_{e^+e^-\rightarrow q X}/d^3p_1$.
Since $x_1<1$ and $\hat{y}_{13}\geq\bar{y}_m > 0$, there is neither an
infrared nor a collinear divergence in this region.  We perform  
the integration of Eq.~(\ref{e46}) over $\hat{y}_{13}$ in $n=4$
dimensions, and we obtain
\begin{eqnarray}
E_1
\frac{d\hat{\sigma}^{(1)sub}_{e^+e^-\rightarrow qX}}{d^3p_1}
&= &\, \left[\frac{2}{s}F^{PC}_{q}(s)\right]
               \left[\alpha_{em}^2N_c\ \frac{1}{x_1} \right]
               C_F \left(\frac{\alpha_s}{2\pi}\right)  \nonumber \\
&& \times (1+\cos^2\theta_1)
         \left[\left(\frac{1+x_1^2}{1-x_1}\right)
               \ell n \left( 
                 \frac{(1-x_1)\delta^2/4}
                      {(1+\epsilon_h)(x_\gamma/x_1)-1/x_1} 
               \right) \right]\ .
\label{e54}
\end{eqnarray}
Terms of $O(\delta^2)$ are dropped.  In deriving Eq.~(\ref{e54}), we
used the fact that the second term in Eq.~(\ref{e19}) vanishes in this 
region.   

To summarize the derivation of this subsection, 
we present a general form for the $O(\alpha_s)$ subtraction term to 
$\sigma^{(1)sub}_{e^+e^-\rightarrow\gamma X}$ via quark fragmentation, 
\begin{equation}
E_\gamma 
\frac{d\sigma^{(1)sub}_{e^+e^-\rightarrow qX\rightarrow \gamma X} }
     {d^3\ell}
 = \sum_q
   \int^1_{\max\left[x_\gamma,\frac{1}{1+\epsilon_h}\right]}\, 
   \frac{dz}{z}
\left[ E_1 \frac{d\hat{\sigma}_{e^+e^-\rightarrow qX}^{(1)sub}}
                {d^3p_1}
           \left(x_1=\frac{x_\gamma}{z}\right) \right]
\frac{D_{q\rightarrow\gamma}(z, \mu^2_F)}{z}\ .
\label{e54b}
\end{equation}
For collinear fragmentation $q\rightarrow \gamma$, $\theta_3 = 
\theta_\gamma$.

We remark again, as we did in our discussions of the direct and gluon 
fragmentation contributions, that after 
convolution with the quark-to-photon fragmentation function, 
the subtraction term in Eq.~(\ref{e54b}) manifests
a logarithmic divergence of the form 
$\ell n(1+\epsilon_h-1/x_\gamma)$ when $x_\gamma \rightarrow
1/(1+\epsilon_h)$. 
%%%%%%%%%%%%%% End of Section III %%%%%%%%%%%%%%%%%%%%%%%%%%%%%%%%%%%%%

%%%%%%%%%%%%%% Begin Section IV %%%%%%%%%%%%%%%%%%%%%%%%%%%%%%%%%%%%%%%
\section{One-Loop Contributions to the Isolated Cross Section}
\label{sec:4}

Combining the one-loop subtraction terms calculated in last section
and the one-loop contributions to the inclusive cross section derived
in Ref.~\cite{BGQ1}, we present in this section the complete one-loop
contributions to the cross section for isolated photons in $e^+e^-$
collisions. 

For the convenience of later discussion, we first list all one-loop
contributions to the inclusive cross section, and then we present 
analytical results for the isolated cross section in  
three subsections corresponding to $x_\gamma < 1/(1+\epsilon_h)$, 
$x_\gamma > 1/(1+\epsilon_h)$, and $x_\gamma = 1/(1+\epsilon_h)$, 
respectively.  We conclude this section with a discussion of the 
logarithmic divergence associated with the special point $x_\gamma
\Rightarrow 1/(1+\epsilon_h)$.

\subsection{One-Loop Contributions to the Inclusive Cross Section}
\label{subsec:4a}

Complete one-loop contributions to the inclusive cross section, 
$E_\gamma d\sigma^{incl}_{e^+e^-\rightarrow\gamma X}/d^3\ell$, 
are derived in Ref.~\cite{BGQ1}.  The results are summarized as 
follows.

\noindent {\it One-loop direct production of $e^+e^-\rightarrow\gamma$
at $O(\alpha_{em})$:}
\begin{eqnarray}
E_\gamma
\frac{d\hat{\sigma}^{(1)incl}_{e^+e^-\rightarrow\gamma X}}{d^3\ell}
&= & \, 2 \sum_{q} \left[\frac{2}{s}F^{PC}_{q}(s)\right]
               \left[\alpha_{em}^2N_c \frac{1}{x_\gamma} \right]
          e_q^2 \left(\frac{\alpha_{em}}{2\pi}\right) \nonumber \\
&\times & \Bigg\{(1+\cos^2\theta_\gamma)
               \left(\frac{1+(1-x_\gamma)^2}{x_\gamma}\right)
               \left[\ell n\left(\frac{s}{\mu^2_{\overline{\rm MS}}}\right)
                         + \ell n(x_\gamma^2(1-x_\gamma)) \right]
    \nonumber \\
&& \quad 
+ (1-3\cos^2\theta_\gamma)\left[\frac{2(1-x_\gamma)}{x_\gamma}\right]
\Bigg\}\ .
\label{e58}
\end{eqnarray}

\noindent {\it One-loop fragmentation contribution via a gluon: }
$e^+e^-\rightarrow g \rightarrow \gamma$ at 
$O(\alpha_{s})$:
\begin{equation}
E_\gamma {{d\sigma^{(1)incl}_{e^+e^-\rightarrow gX\rightarrow \gamma X}} 
\over {d^3\ell}} = \int^1_{x_\gamma}\, {{dz} \over {z}} 
\left[ E_g {{d\hat{\sigma}_{e^+e^-\rightarrow gX}^{(1)incl}}\over{d^3p_g}}\,
\left( x_g = {{x_\gamma} \over {z}}\right)\right]\, 
{{D_{g\rightarrow\gamma}(z, \mu^2_F)} \over {z}} ,
\label{e59}
\end{equation}
where the short-distance hard part is
\begin{eqnarray}
E_g \frac{d\hat{\sigma}^{(1)incl}_{e^+e^-\rightarrow gX}}{d^3p_g}
&= & \, 2 \sum_{q} \left[\frac{2}{s}F^{PC}_{q}(s)\right]
               \left[\alpha_{em}^2N_c \frac{1}{x_g} \right]
           C_F \left(\frac{\alpha_{s}}{2\pi}\right) \nonumber \\
&\times & \Bigg\{(1+\cos^2\theta_\gamma)
               \left(\frac{1+(1-x_g)^2}{x_g}\right)
               \left[\ell n\left(\frac{s}{\mu^2_{\overline{\rm MS}}}\right)
                         + \ell n(x_g^2(1-x_g)) \right]\nonumber \\
&& \quad 
+ (1-3\cos^2\theta_\gamma)\left[\frac{2(1-x_g)}{x_g}\right]
\Bigg\},
\label{e60}
\end{eqnarray}
with $x_g = 2E_g/\sqrt{s}$, and $\theta_g=\theta_\gamma$.  

\noindent {\it One-loop fragmentation contribution through a quark, 
$e^+e^-\rightarrow q \rightarrow \gamma$ at 
$O(\alpha_{s})$:}
\begin{equation}
E_\gamma {{d\sigma^{(1)incl}_{e^+e^-\rightarrow qX\rightarrow \gamma X}} 
\over {d^3\ell}} = \sum_{q} \int^1_{x_\gamma}\, {{dz} \over {z}} 
\left[ 
 E_1 {{d\hat{\sigma}_{e^+e^-\rightarrow qX}^{(1)incl}}\over{d^3p_1}}\,
 \left( x_1 = {{x_\gamma} \over {z}}\right)\right]\, 
{{D_{q\rightarrow\gamma}(z, \mu^2_F)} \over {z}}\ ,
\label{e61}
\end{equation}
where the short-distance hard part is
\begin{eqnarray}
E_1\frac{d\hat{\sigma}^{(1)incl}_{e^+e^-\rightarrow qX}}{d^3p_1}
&= & \, \left[\frac{2}{s}F^{PC}_{q}(s)\right]
               \left[\alpha_{em}^2N_c \frac{1}{x_1} \right]
           C_F \left(\frac{\alpha_{s}}{2\pi}\right) \nonumber \\
&\times & \Bigg\{(1+\cos^2\theta_\gamma)
   \Bigg[\left(\frac{1+x_1^2}{(1-x_1)_+}+\frac{3}{2}\delta(1-x_1)\right)
               \ell n\left(\frac{s}{\mu^2_{\overline{\rm MS}}}\right) 
        \nonumber \\
&&\quad\quad + \left(\frac{1+x_1^2}{1-x_1}\right)\ell n\left(x_1^2\right)
       + \left(1+x_1^2\right)\left(\frac{\ell n(1-x_1)}{1-x_1}\right)_+
       - \frac{3}{2}\left(\frac{1}{1-x_1}\right)_+ \nonumber \\
&&\quad\quad + \delta(1-x_1)\left(\frac{2\pi^2}{3}-\frac{9}{2}\right)
       - \frac{1}{2}\left(3x_1-5\right) \Bigg] \nonumber \\
&&\quad + \left(1-3\cos^2\theta_\gamma\right) \Bigg\} \ .
\label{e62}
\end{eqnarray}
Here $\theta_1=\theta_\gamma$ is used, and the ``+'' description is 
defined in Eq.~(\ref{e20a}).

\subsection{One-Loop Contribution to the Isolated Cross Section 
when $x_\gamma < 1/(1+\epsilon_h)$}
\label{subsec:4b}

Substituting Eqs.~(\ref{e36}) and (\ref{e58}) into Eq.~(\ref{e10a}) for 
$c=\gamma$, we derive the $O(\alpha_{em})$ one-loop direct production
of isolated photons via $e^+e^-\rightarrow\gamma$:
\begin{eqnarray}
E_\gamma
\frac{d\hat{\sigma}^{(1)iso}_{e^+e^-\rightarrow\gamma X}}{d^3\ell}
&= & \, 2 \sum_{q} \left[\frac{2}{s}F^{PC}_{q}(s)\right]
               \left[\alpha_{em}^2N_c \frac{1}{x_\gamma} \right]
          e_q^2 \left(\frac{\alpha_{em}}{2\pi}\right) \nonumber \\
&\times & \Bigg\{(1+\cos^2\theta_\gamma)
         \left[\left(\frac{1+(1-x_\gamma)^2}{x_\gamma}\right)
               \ell n\left(\frac{1}{(1-x_\gamma)\delta^2/4}\right)
               -x_\gamma  \right]\nonumber \\
&& 
+ (1-3\cos^2\theta_\gamma)\left[\frac{2(1-x_\gamma)}{x_\gamma}\right]
\Bigg\}\ ,
\label{e63}
\end{eqnarray}
where we have dropped terms of $O(\delta^2)$.  By integrating over 
$\theta_\gamma$, we may verify that Eq.~(\ref{e63}) is consistent with
the analogous expression derived in Ref.~\cite{KT}.

From Eqs.~(\ref{e9a}), (\ref{e37}) and (\ref{e60}), we obtain the
$O(\alpha_s)$ one-loop gluonic fragmentation contribution to the cross
section for isolated photons:   
\begin{equation}
E_\gamma 
\frac{d\sigma^{(1)iso}_{e^+e^-\rightarrow gX\rightarrow \gamma X}} 
     {d^3\ell} = 
     \int^1_{\max\left[x_\gamma,\frac{1}{1+\epsilon_h}\right]}\, 
     \frac{dz}{z} \left[ 
 E_g \frac{d\hat{\sigma}_{e^+e^-\rightarrow gX}^{(1)iso}}{d^3p_g}
\left( x_g = \frac{x_\gamma}{z} \right) \right] 
\frac{D_{g\rightarrow\gamma}(z, \mu^2_F)}{z}\ ,
\label{e64}
\end{equation}
where the short-distance hard part is 
\begin{eqnarray}
E_g
\frac{d\hat{\sigma}^{(1)iso}_{e^+e^-\rightarrow gX}}{d^3p_g}
&= & \, 2 \sum_{q} \left[\frac{2}{s}F^{PC}_{q}(s)\right]
               \left[\alpha_{em}^2N_c \frac{1}{x_g} \right]
          C_F\, \left(\frac{\alpha_{s}}{2\pi}\right) \nonumber \\
&\times & \Bigg\{(1+\cos^2\theta_\gamma)
         \left[\left(\frac{1+(1-x_g)^2}{x_g}\right)
               \ell n\left(\frac{1}{(1-x_g)\delta^2/4}\right)
               -x_g  \right]\nonumber \\
&& 
+ (1-3\cos^2\theta_\gamma)\left[\frac{2(1-x_g)}{x_g}\right]
\Bigg\}\ .
\label{e65}
\end{eqnarray}

Similarly, we derive the $O(\alpha_s)$ one-loop quark fragmentation
contribution to the cross section for isolated photons: 
\begin{equation}
E_\gamma 
\frac{d\sigma^{(1)iso}_{e^+e^-\rightarrow qX\rightarrow \gamma X}} 
     {d^3\ell} = \sum_{q}
\int^1_{\max\left[x_\gamma,\frac{1}{1+\epsilon_h}\right]} 
\frac{dz}{z} \left[ 
E_1 \frac{d\hat{\sigma}_{e^+e^-\rightarrow qX}^{(1)iso}}{d^3p_1}
\left( x_1 = \frac{x_\gamma}{z}\right)\right] 
\frac{D_{q\rightarrow\gamma}(z, \mu^2_F)}{z}\ ,
\label{e66}
\end{equation}
where the short-distance hard part is
\begin{eqnarray}
E_q
\frac{d\hat{\sigma}^{(1)iso}_{e^+e^-\rightarrow qX}}{d^3p_q}
&= & \, \left[\frac{2}{s}F^{PC}_{q}(s)\right]
         \left[\alpha_{em}^2N_c \frac{1}{x_1} \right]
          C_F\, \left(\frac{\alpha_{s}}{2\pi}\right) \nonumber \\
&\times & \Bigg\{\left(1+\cos^2\theta_\gamma\right)
         \left[\left(\frac{1+x_1^2}{1-x_1}\right)
               \ell n\left(\frac{1}{(1-x_1)\delta^2/4}\right)
               -\frac{3}{2}\left(\frac{1}{1-x_1} \right)
               -\frac{1}{2}\left(x_1-3\right) \right]\nonumber \\
&& 
+ \left(1-3\cos^2\theta_\gamma\right) \Bigg\}\ .
\label{e67}
\end{eqnarray}
Equation (\ref{e67}) is derived by substituting 
Eqs.~(\ref{e52}) and (\ref{e62}) into Eq.~(\ref{e10b}).  We observe
after integrating over $z$, that Eq.~(\ref{e66}) develops a 
logarithmic divergence $\ell n(1/x_\gamma-(1+\epsilon_h))$ as
$x_\gamma\rightarrow 1/(1+\epsilon_h)$.  This divergence is caused by
the $1/(1-x_1)$ terms in Eq.~(\ref{e67}).  A detailed discussion of
this divergence is presented later in subsection~\ref{subsec:4d}.

Assuming $D_{q\rightarrow\gamma}(z)
=D_{\bar{q}\rightarrow\gamma}(z)$, 
the one-loop antiquark fragmentation contribution to the cross section 
for isolated photons is the same as that of the quark contribution in
Eq.~(\ref{e66}), except that the $\sum$ is over $\bar{q}$.   

\subsection{One-Loop Contribution to the Isolated Cross Section
when $x_\gamma > 1/(1+\epsilon_h)$}
\label{subsec:4c}

As pointed out in Section~\ref{sec:3}, the subtraction terms 
$\hat{\sigma}^{(1)sub}$ would vanish for $x_\gamma > 1/(1+\epsilon_h)$ 
if all fragmentation processes were exactly collinear.  But, due to
the finite cone size, there is a small region of phase space where the
subtraction terms are finite.  Since the value of $\epsilon_h$ is very
small in most experiments, very limited data from $e^+e^-\rightarrow
\gamma X$ are available in this region.  Nevertheless, the isolated 
cross section in this region is interesting theoretically and
important for understanding the isolated cross section in
hadron-hadron collisions. 

Kinematics require the subtraction term $E_\gamma 
d\hat{\sigma}^{(1)sub}_{e^+e^-\rightarrow\gamma X}/ d^3\ell$ to vanish
if $x_\gamma > x_\gamma^{\rm max}(\delta,\epsilon_h)$, defined 
in Eq.~(\ref{e32a}); therefore, 
\begin{equation}
E_\gamma 
\frac{d\hat{\sigma}^{(1)iso}_{e^+e^-\rightarrow\gamma X}}{d^3\ell}
= E_\gamma 
\frac{d\hat{\sigma}^{(1)incl}_{e^+e^-\rightarrow\gamma X}}{d^3\ell}
\quad\quad \mbox{if}\
x_\gamma > x_\gamma^{\rm max}(\delta,\epsilon_h) .
\label{e69}
\end{equation}
However, if $1/(1+\epsilon_h) < x_\gamma < 
x_\gamma^{\rm max}(\delta,\epsilon_h)$, we use Eqs.~(\ref{e34}) and
(\ref{e58}) to derive
\begin{eqnarray}
E_\gamma
\frac{d\hat{\sigma}^{(1)iso}_{e^+e^-\rightarrow\gamma X}}{d^3\ell}
&= & \, 2 \sum_{q} \left[\frac{2}{s}F^{PC}_{q}(s)\right]
               \left[\alpha_{em}^2N_c \frac{1}{x_\gamma} \right]
          e_q^2 \left(\frac{\alpha_{em}}{2\pi}\right) \nonumber \\
&\times & \Bigg\{\left(1+\cos^2\theta_\gamma\right)
               \left(\frac{1+(1-x_\gamma)^2}{x_\gamma}\right)
               \Bigg[\ell n\left(\frac{s}{\mu^2_{\overline{\rm MS}}}\right)
                         + \ell n(x_\gamma^2(1-x_\gamma)) \nonumber \\
&& \mbox{\hskip 2.1 in} + \ell n\left(
     \frac{1+\epsilon_h-1/x_\gamma}{(1-x_\gamma)\delta^2/4}\right) 
     \Bigg] \nonumber \\
&& + \left(1-3\cos^2\theta_\gamma\right)
    \left[\frac{2(1-x_\gamma)}{x_\gamma}\right]
\Bigg\} \ .
\label{e70}
\end{eqnarray}
If we integrate Eq.~(\ref{e70}) over $\theta_\gamma$, we obtain
the analogous result derived previously in Ref.~\cite{KT}.

For the one-loop fragmentation contributions from parton
$c(=g,q,\bar{q})$ to $\gamma$, the subtraction terms again vanish if  
$x_c > x^{\rm max}(z,\delta,\epsilon_h)$, defined 
in Eq.~(\ref{e44a}); thus,
\begin{equation}
E_c 
\frac{d\hat{\sigma}^{(1)iso}_{e^+e^-\rightarrow c X}}{d^3p_c}
= E_c 
\frac{d\hat{\sigma}^{(1)incl}_{e^+e^-\rightarrow c X}}{d^3p_c}
\quad\quad \mbox{if}\
x_c > x^{\rm max}(z,\delta,\epsilon_h)\ .
\label{e71}
\end{equation}
In Eq.~(\ref{e71}), the corresponding inclusive hard parts are given in 
Eqs.~(\ref{e60}) and (\ref{e62}).

If $1/(1+\epsilon_h) < x_c < x^{\rm max}(z,\delta,\epsilon_h)$,
the subtraction terms are finite and are
given in Eqs.~(\ref{e38}) and (\ref{e54}).  
Combining Eq.~(\ref{e38}) with the inclusive contribution, we obtain
the gluon fragmentation contribution as
\begin{eqnarray}
E_g \frac{d\hat{\sigma}^{(1)iso}_{e^+e^-\rightarrow gX}}{d^3p_g}
&= & \, 2 \sum_{q} \left[\frac{2}{s}F^{PC}_{q}(s)\right]
               \left[\alpha_{em}^2N_c \frac{1}{x_g} \right]
           C_F \left(\frac{\alpha_{s}}{2\pi}\right) \nonumber \\
&\times & \Bigg\{(1+\cos^2\theta_\gamma)
               \left(\frac{1+(1-x_g)^2}{x_g}\right)
               \Bigg[\ell n\left(\frac{s}{\mu^2_{\overline{\rm MS}}}\right)
                         + \ell n(x_g^2(1-x_g)) \nonumber \\
&& \mbox{\hskip 2.1 in} + \ell n \left( 
                 \frac{(1+\epsilon_h)(x_\gamma/x_g)-1/x_g}
                      {(1-x_g)\delta^2/4} \right) \Bigg] \nonumber \\
&& 
+ (1-3\cos^2\theta_\gamma)\left[\frac{2(1-x_g)}{x_g}\right]
\Bigg\}
\label{e72}
\end{eqnarray}
for $1/(1+\epsilon_h) < x_g < x^{\rm max}(z,\delta,\epsilon_h)$. 
The quark fragmentation contribution is
\begin{eqnarray}
E_q \frac{d\hat{\sigma}^{(1)iso}_{e^+e^-\rightarrow qX}}{d^3p_q}
&= & \, \left[\frac{2}{s}F^{PC}_{q}(s)\right]
               \left[\alpha_{em}^2N_c \frac{1}{x_1} \right]
           C_F \left(\frac{\alpha_{s}}{2\pi}\right) \nonumber \\
&\times & \Bigg\{(1+\cos^2\theta_\gamma)
   \Bigg[\left(\frac{1+x_1^2}{1-x_1}\right)\left[
               \ell n\left(\frac{s}{\mu^2_{\overline{\rm MS}}}\right)
             + \ell n\left(x_1^2(1-x_1)\right) \right. \nonumber \\
&&\quad\quad \left.
             + \ell n\left(\frac{(1+\epsilon_h)(x_\gamma/x_1)-1/x_1}
                                {(1-x_1)\delta^2/4}\right) \right]
             - \frac{3}{2}\left(\frac{1}{1-x_1}\right)
             - \frac{1}{2}\left(3x_1-5\right) \Bigg] \nonumber \\
&& + \left(1-3\cos^2\theta_\gamma\right) \Bigg\} 
\label{e73}
\end{eqnarray}
for $1/(1+\epsilon_h) < x_1 < x^{\rm max}(z,\delta,\epsilon_h)$; 
$x^{\rm max}(z,\delta,\epsilon_h)$ is defined in Eq.~(\ref{e44a}).

It is a common feature of Eqs.~(\ref{e70}), (\ref{e72}) and
(\ref{e73}) that all three develop a logarithmic divergence 
$\ell n(1+\epsilon_h - 1/x_\gamma)$ as 
$x_\gamma\rightarrow 1/(1+\epsilon_h)$.  This logarithm results from
the collinear singularities of the Feynman diagrams shown in 
Figs.~\ref{fig4}, \ref{fig6} and \ref{fig7}a.  
It corresponds to the situation in which an unobserved
parton inside the isolation cone becomes almost collinear with the
observed parton.  Since $x_\gamma > 1/(1+\epsilon_h)$,  
which is the same as $x_g$ or $x_q > 1/(1+\epsilon_{\min}(z))$ for all
$z$, the counter terms, defined through $\ddot{\otimes}$ 
in Eqs.~(\ref{e16}), (\ref{e18}), and (\ref{e19}), vanish.
Consequently, the collinear singularities have no corresponding
canceling counter terms.

This problem is caused by the incompatibility between collinear 
fragmentation, which was used to define the counter terms in 
Eqs.~(\ref{e16}), (\ref{e18}) and (\ref{e19}),
and the cone fragmentation used to define the partonic cross section
$E_\gamma d\hat{\sigma}^{(1)sub}_{e^+e^-\rightarrow cX}/d^3\ell$
with $c=\gamma, q, \bar{q}, g$ in Eqs.~(\ref{e16})
(\ref{e18}), and (\ref{e19}).  As we pointed out in
Section~\ref{sec:1}, to deal with the region $x_\gamma >
1/(1+\epsilon_h)$ it may be necessary to revise our concept of photon
fragmentation functions in the case of isolated photons.
  
\subsection{One-Loop Contribution to the Isolated Cross Section
            when $x_\gamma = 1/(1+\epsilon_h)$}
\label{subsec:4d}

As $x_\gamma$ approaches $1/(1+\epsilon_h)$, a number of the
expressions derived above for the isolated cross section develop a
logarithmic divergence of the form $\ell n|(1+\epsilon_h)-1/x_\gamma
|$.  However, the value of $\epsilon_h$ is an arbitrary parameter
chosen in individual experiments.  Certainly, a completely consistent
theoretical prediction should not  
be sensitive to an arbitrary experimental parameter, such as 
$\epsilon_h$.  We argue below that the reason for the 
unstable result as $x_\gamma$ approaches $1/(1+\epsilon_h)$ is 
the breakdown \cite{BGQ2} of the conventional perturbative
factorization theorem for the cross section of isolated photons in the
neighborhood of $x_\gamma=1/(1+\epsilon_h)$.

In this subsection, we first use our results of the last two
subsections to explain the origin of this logarithmic divergence.
Then, we present a derivation of the isolated cross section 
for $x_\gamma$ near the value $1/(1+\epsilon_h)$, and we show that the
key issues are the isolation condition and the finite cone size for
fragmentation. 

We start with the situation in which $x_\gamma$ approaches 
$1/(1+\epsilon_h)$ from below.  When $x_\gamma$ is less than 
$1/(1+\epsilon_h)$, the expression for direct production, 
given in Eq.~(\ref{e63}), is well-behaved as 
$x_\gamma$ approaches $1/(1+\epsilon_h)$.  Similarly,
the gluonic fragmentation contribution, defined by Eqs.~(\ref{e64}) and 
(\ref{e65}), is also well-behaved as 
$x_\gamma$ approaches $1/(1+\epsilon_h)$.  Although $x_g$ can 
equal 1 when $x_\gamma=1/(1+\epsilon_h)$, the $\ell n(1-x_g)$ term in 
Eq.~(\ref{e65}) gives a finite contribution after the integration over
$z$.  Thus, Eqs.~(\ref{e63}) and (\ref{e64}) should also be valid at
$x_\gamma = 1/(1+\epsilon_h)$.

On the other hand, the quark (or antiquark) fragmentation contribution
develops a logarithmic divergence as $x_\gamma$ approaches 
$1/(1+\epsilon_h)$.  This divergence is caused by 
the $1/(1-x_1)$ terms in Eq.~(\ref{e67}), and it can be 
understood as follows.  Consider the following general integral
\begin{equation}
I(x_\gamma,\epsilon_h) \equiv 
\int_{1/(1+\epsilon_h)}^1 \frac{dz}{z} \left(\frac{1}{1-x_1}\right)
\ell n^m(1-x_1) F(z,x_1=x_\gamma/z) ,
\quad\quad \mbox{for}\ 
x_\gamma \leq 1/(1+\epsilon_h) ,
\label{e74}
\end{equation}
where $m=0,1,...$, and $F(z,x_1=x_\gamma/z)$ is any smooth function
over the region of integration.  The integral $I(x_\gamma,\epsilon_h)$
can be thought of as a simplified version of the one-loop quark
fragmentation contribution defined in Eq.~(\ref{e66}).  The factors
$1/(1-x_1)$ and $\ell n(1-x_1)$ are typical of the short-distance hard
part.   It is straightforward to perform the integration, and we find 
\begin{eqnarray}
I(x_\gamma,\epsilon_h) 
&=&  \int_{x_\gamma}^{x_\gamma(1+\epsilon_h)} \frac{dx_1}{x_1} 
    \left(\frac{1}{1-x_1}\right)
    \ell n^m(1-x_1) F(z=x_\gamma/x_1,x_1) \nonumber \\
&\Rightarrow & -\ell n^{m+1}\left(1-x_\gamma(1+\epsilon_h)\right)
\quad \Rightarrow \pm\infty 
\quad\quad \mbox{as}\
           x_\gamma \rightarrow 1/(1+\epsilon_h).
\label{e75}
\end{eqnarray}
This example shows that $1/(1-x_1)$ terms in the 
short-distance hard part make the isolated cross section very 
sensitive to the value of $\epsilon_h$, as $x_\gamma$ approaches 
$1/(1+\epsilon_h)$.  The perturbatively calculated 
cross section for isolated photons becomes logarithmically divergent
at a different value of $x_\gamma$ if one chooses 
a different value of $\epsilon_h$.  

The reason for this unsatisfactory logarithmic sensitivity is 
the infrared divergence associated with the limit in which the
final-state gluon's momentum goes to zero, which is the same as
$x_1\rightarrow 1$.  Normally, such an infrared divergence is
canceled by virtual diagrams.  For example, in the calculation of the
inclusive contribution, the infrared singularity associated with
$x_1\rightarrow 1$ from the real gluon emission diagrams, sketched in
Fig.~\ref{fig7}a, is canceled by the infrared contribution from the
virtual gluon exchange diagrams, sketched in Fig.~\ref{fig7}b,
proportional to $\delta(1-x_1)$.  However, as we demonstrate below, 
perfect cancellation between real and virtual diagrams is upset 
by the isolation conditions.

Isolated photons are defined to be photons accompanied by less than
$E_{max}=\epsilon_h E_\gamma$ of hadronic energy in the isolation
cone.  For photons produced from parton fragmentation, as sketched in
Fig.~\ref{fig2}, a situation can arise in which $E_{max}$ in the
isolation cone is provided completely by parton fragmentation.
Consequently, no other soft gluons are allowed to enter the isolation
cone.  The phase space for the final-state non-fragmenting partons is
reduced by the size of the cone.  Due to the mismatch of the phase
space, the infrared divergences associated with the final-state real
gluons cannot be completely canceled by the virtual diagrams.
Such uncanceled infrared contributions vanish if the cone size is
zero. 

As an example, we use the one-loop quark fragmentation contribution to
the cross section for isolated photons to demonstrate the breakdown of
the perfect cancellation of infrared divergences.  In order to make
our presentation parallel the statements of the last paragraph, we
calculate the isolated cross section from the quark fragmentation term
directly without going through the subtraction term \cite{BGQ2}.

To examine the situation in which $x_\gamma$ approaches
$1/(1+\epsilon_h)$, we must evaluate both the real and the virtual
diagrams shown in Fig.~\ref{fig7}.  The squared matrix element for the
real diagrams is obtained from Eq.~(\ref{e46}):
\begin{eqnarray}
\frac{1}{4}H^{real}
&=&\ \frac{1}{4}\left( H_1+H_2^{eff} \right) \nonumber \\
&=&\ (e\mu^\epsilon)^2 (g\mu^\epsilon)^2 
\Bigg\{ 
 \left( 1 + \cos^2 \theta_1 - 2\epsilon\right)
  \left[ \left( \frac{1+x_1^2}{1-x_1} \right) \frac{1}{y_{13}} 
  + \frac{y_{13}}{1-x_1} \right] \nonumber \\
&&+ \left( 1 + \cos^2\theta_1 - 2\epsilon\right) 
  \left[ -\frac{2}{1-x_1}
         -\epsilon\left( \frac{1-x_1}{{y}_{13}}+
                         \frac{{y}_{13}}{1-x_1}+2 \right) \right]
         \nonumber \\
&&+ \left( 1-3\cos^2\theta_1\right) 
  \left[ \frac{2}{x_1} \left(1-\frac{y_{13}}{x_1}\right) \right] 
  \Bigg\}\ .
\label{e88}
\end{eqnarray}
Equation (\ref{e88}) holds for both the inclusive and the isolated
cross sections.  The key difference resides in the limits of
integration over $\hat{y}_{13} = 
y_{13}/x_1$.  Integrating $\hat{y}_{13}$ from 0 to 1, we 
obtain the real contribution to the inclusive cross section; all
infrared divergences associated with $x_1\rightarrow 1$ are canceled
by the contribution from the virtual diagrams \cite{BGQ1}.  On the
other hand, the isolation conditions require that the $\hat{y}_{13}$
integration be divided into three regions:
\begin{equation}
\int d\hat{y}_{13} \Rightarrow 
 \int_{0}^{\min[\bar{y}_c,\bar{y}_m]} d\hat{y}_{13} + 
 \int_{\bar{y}_c}^{1-\bar{y}_c} d\hat{y}_{13} + 
 \int_{\max[(1-\bar{y}_c),(1-\bar{y}_m)]}^{1} d\hat{y}_{13}\ ,
\label{e89}
\end{equation}
where $\bar{y}_c$ and $\bar{y}_m$ are defined in Eq.~(\ref{e48}).
In the first region, the condition $0\leq \hat{y}_{13} \leq \bar{y}_c$
ensures that a gluon is in the isolation cone of the fragmenting
quark; and  
condition $\hat{y}_{13}\leq \min[\bar{y}_c,\bar{y}_m]$ ensures that
the total hadronic energy in the isolation cone is less than $E_{max}=
\epsilon_h E_\gamma$. Similarly, the condition $\max[(1-\bar{y}_c),
(1-\bar{y}_m)] \leq \hat{y}_{13} \leq 1$ for the third region ensures
that the antiquark is in the isolation cone and that the total
hadronic energy in the isolation cone is less than $E_{max}$.  The
second interval represents the situation in which neither the gluon
nor the antiquark is in the isolation cone.  

If $\bar{y}_c \leq \bar{y}_m$, Eq.~({\ref{e89}) shows that the
isolated cross section is the same as the inclusive cross section,
consistent with Eq.~(\ref{e49}).  When $\bar{y}_c > 
\bar{y}_m$, (i.e., $(1-x_1)\delta^2/4 >
z[(1+\epsilon_h)-1/x_\gamma]$), Eq.~(\ref{e89}) states that 
the phase space of the final state gluon (and/or antiquark) 
is \underline{smaller} than in the case of the inclusive
cross section.  

If $x_\gamma \leq 1/(1+\epsilon_h)$, or, equivalently, $\bar{y}_m =
z[(1+\epsilon_h)-1/x_\gamma] \leq 0$, the
$\hat{y}_{13}$ integration defined in Eq.~(\ref{e89}) reduces to 
the second region only 
\begin{equation}
\int d\hat{y}_{13} \Rightarrow 
 \int_{(1-x_1)\delta^2/4}^{1-(1-x_1)\delta^2/4} d\hat{y}_{13},
\label{e90}
\end{equation}
where we expand $\bar{y}_c$ to order $\delta^2$.  We rewrite the
limits of the integration over $\hat{y}_{13}$ in Eq.~(\ref{e90}) 
as 
\begin{equation}
 \int_{(1-x_1)\delta^2/4}^{1-(1-x_1)\delta^2/4} d\hat{y}_{13} = 
 \int_0^1 d\hat{y}_{13} -
 \int_0^{(1-x_1)\delta^2/4} d\hat{y}_{13} -
 \int_{1-(1-x_1)\delta^2/4}^1 d\hat{y}_{13} .
\label{e91}
\end{equation}
Combining the three-particle phase space of Eq.~(\ref{e23})
and the squared matrix element of Eq.~(\ref{e88}), we derive that the 
\underline{first} term on the right-hand-side of Eq.~(\ref{e91})
provides the complete real gluon emission contribution for the
inclusive cross section \cite{BGQ1} 
\begin{eqnarray}
E_1 \frac{d\sigma^{(R_1)}_{e^+e^-\rightarrow qX}}{d^3p_1} 
&=& \left[ \frac{2}{s}\, F^{PC} (s)\right] 
   \left[ \alpha^2_{em}\, N_c 
   \left( \frac{4\pi \mu^2}{(s/4) \sin^2\theta_1}\right)^\epsilon 
          \frac{1}{\Gamma(1-\epsilon)}\, \frac{1}{x_1} \right]
\nonumber \\
&\times &C_F \left( \frac{\alpha_s}{2\pi}\right)
\left[ \left(\frac{4\pi \mu^2}{s}\right)^\epsilon
             \frac{1}{\Gamma(1-\epsilon)} \right]\,
             \frac{\Gamma(1-\epsilon)^2}{\Gamma(1-2\epsilon)} 
\nonumber \\
&\times &\Bigg\{
 \left( 1 + \cos^2 \theta_1 - 2\epsilon\right)\Bigg[
      \left(\frac{1+x_1^2}{(1-x_1)_+}+\frac{3}{2}\delta(1-x_1)\right)
      \left(\frac{1}{-\epsilon}\right) \nonumber \\
&&\quad\quad\quad 
       +\left(\frac{1+x_1^2}{1-x_1}\right)\ell n\left(x_1^2\right)
       +\left(1+x_1^2\right)\left(\frac{\ell n(1-x_1)}{1-x_1}\right)_+
       -\frac{3}{2}\left(\frac{1}{1-x_1}\right)_+ \nonumber \\
&&\quad\quad\quad 
       +\delta(1-x_1)\left(
       \frac{2}{\epsilon^2}+\frac{3}{\epsilon}+\frac{7}{2} \right)
      -\frac{1}{2}\left(3x_1-5\right) \Bigg] \nonumber \\
&+& \left(1-3\cos^2\theta_1 \right) \Bigg\} \ .
\label{realg}
\end{eqnarray}
The superscript $(R_1)$ stands for the contribution of real
gluon emission from the \underline{first}
term on the right-hand-side of Eq.~(\ref{e91}).  In deriving
Eq.~(\ref{realg}), we use 
\begin{equation}
\frac{1}{(1-x_1)^{1+\epsilon}}=
-\frac{1}{\epsilon}\delta(1-x_1)
+\left(\frac{1}{1-x_1}\right)_+
-\epsilon \left(\frac{\ell n(1-x_1)}{1-x_1}\right)_+ .
\label{plus2e}
\end{equation}
The virtual gluon exchange contribution from the diagrams in
Fig.~\ref{fig7}b is \cite{BGQ1}
\begin{eqnarray}
E_1 \frac{d\sigma^{(V)}_{e^+e^-\rightarrow qX}}{d^3p_1} 
&=& \left[ \frac{2}{s}\, F^{PC} (s)\right] 
   \left[ \alpha^2_{em}\, N_c 
   \left( \frac{4\pi \mu^2}{(s/4) \sin^2\theta_1}\right)^\epsilon 
          \frac{1}{\Gamma(1-\epsilon)}\, \frac{1}{x_1} \right]
\nonumber \\
&\times &C_F \left( \frac{\alpha_s}{2\pi}\right)
\left[ \left(\frac{4\pi \mu^2}{s}\right)^\epsilon
             \frac{1}{\Gamma(1-\epsilon)} \right]\,
             \frac{\Gamma(1-\epsilon)^3\Gamma(1+\epsilon)}
                  {\Gamma(1-2\epsilon)} \nonumber \\
&\times &\Bigg\{
 \left( 1 + \cos^2 \theta_1 - 2\epsilon\right)
       \delta(1-x_1)
       \left[-\frac{2}{\epsilon^2}-\frac{3}{\epsilon}
       +\left(\pi^2-8\right)\right] \Bigg\} \ ,
\label{virtual}
\end{eqnarray}
where the superscript $(V)$ stands for the virtual gluon exchange
contribution. 

Combining the virtual $(V)$ contribution and a part of the real 
contribution $(R_1)$, we derive
\begin{eqnarray}
E_1 \frac{d\sigma^{(R_1+V)}_{e^+e^-\rightarrow qX}}{d^3p_1} 
&=&\ \left[ \frac{2}{s}\, F^{PC} (s)\right] 
    \left[\alpha^2_{em}\, N_c 
    \left(\frac{4\pi \mu^2}{(s/4) \sin^2\theta_1}\right)^\epsilon 
          \frac{1}{\Gamma(1-\epsilon)} \frac{1}{x_1} \right]
\nonumber \\
&&\times C_F \left( \frac{\alpha_s(\mu^2)}{2\pi}\right)
 \Bigg\{
      \left( 1 + \cos^2 \theta_1 - 2\epsilon\right)
      \left[\frac{1+x_1^2}{(1-x_1)_+}+\frac{3}{2}\delta(1-x_1)\right]
      \left(\frac{1}{-\epsilon}\right) \Bigg\} \nonumber \\
&+&\ \left[ \frac{2}{s}\, F^{PC} (s)\right] 
   \left[ \alpha^2_{em}\, N_c\, \frac{1}{x_1} \right]
          C_F \left( \frac{\alpha_s(\mu^2)}{2\pi}\right) 
          \nonumber \\
&&\times\Bigg\{ (1+\cos^2\theta_1)
 \Bigg[\left(\frac{1+x_1^2}{(1-x_1)_+}+\frac{3}{2}\delta(1-x_1)\right)
             \ell n\left(\frac{s}{\mu^2_{\overline{\rm MS}}}\right) 
\nonumber \\
&&\quad\quad 
   + \left(\frac{1+x_1^2}{1-x_1}\right)\ell n\left(x_1^2\right)
   + \left(1+x_1^2\right)\left(\frac{\ell n(1-x_1)}{1-x_1}\right)_+
   - \frac{3}{2}\left(\frac{1}{1-x_1}\right)_+ \nonumber \\
&&\quad\quad 
   + \delta(1-x_1)\left(\frac{2\pi^2}{3}-\frac{9}{2}\right)
   - \frac{1}{2}\left(3x_1-5\right) \Bigg] \nonumber \\
&& + \left(1-3\cos^2\theta_1\right) \Bigg\} \ .
\label{r1v}
\end{eqnarray}
As expected, Eq.~(\ref{r1v}) shows that, except for the $1/\epsilon$
term associated with the collinear singularity between the fragmenting
quark and the real gluon, the contribution from the 
\underline{first} term on the right-hand-side of Eq.~(\ref{e91})
cancels all infrared singularities from the virtual diagrams in
Fig.~\ref{fig7}b. 

The \underline{third} term in Eq.~(\ref{e91}) does not provide any
terms singular in $1/\epsilon$; it yields some 
finite terms that vanish as $\delta^2\rightarrow 0$,
\begin{equation}
E_1 {{d\sigma^{(R_3)}_{e^+e^-\rightarrow qX}} \over {d^3p_1}} 
= O(\delta^2) \ .
\label{r3}
\end{equation}
However, the \underline{second} term generates a number of terms with
$1/\epsilon$ poles.  Neglecting all terms of $O(\delta^2)$ or higher,
we obtain   
\begin{eqnarray}
E_1 \frac{d\sigma^{(R_2)}_{e^+e^-\rightarrow qX}}{d^3p_1}
&=& - \left[ \frac{2}{s}\, F^{PC} (s)\right] 
    \left[ \alpha^2_{em}\, N_c 
    \left( \frac{4\pi \mu^2}{(s/4) \sin^2\theta_1}\right)^\epsilon 
          \frac{1}{\Gamma(1-\epsilon)} \frac{1}{x_1} \right]
\nonumber \\
&&\times C_F \left( \frac{\alpha_s(\mu^2)}{2\pi}\right)
 \Bigg\{
  \left( 1 + \cos^2 \theta_1 - 2\epsilon\right)
  \left[\frac{1+x_1^2}{(1-x_1)_+}+\frac{3}{2}\delta(1-x_1)\right]
  \left(\frac{1}{-\epsilon}\right) \Bigg\} 
\nonumber \\
&-&\ \left[ \frac{2}{s}\, F^{PC} (s)\right] 
   \left[ \alpha^2_{em}\, N_c\, \frac{1}{x_1} \right]
          C_F \left( \frac{\alpha_s(\mu^2)}{2\pi}\right) 
\nonumber \\
&&\times \left(1+\cos^2\theta_1\right)
 \Bigg[
 \left(\frac{1+x_1^2}{(1-x_1)_+}+\frac{3}{2}\delta(1-x_1)\right)
       \ell n\left(\frac{s}{\mu^2_{\overline{\rm MS}}}\right) 
\nonumber \\
&&\quad\quad
    + \frac{1+x_1^2}{1-x_1}\ell n\left(x_1^2\right)
    +2\left(1+x_1^2\right)\left(\frac{\ell n(1-x_1)}{1-x_1}\right)_+
    + \frac{1+x_1^2}{(1-x_1)_+} \ell n\frac{\delta^2}{4}
\nonumber \\
&&\quad\quad
    +\delta(1-x_1)\, \ell n^2\frac{\delta^2}{4}
    +\left(1-x_1\right) \Bigg] 
\nonumber \\
&-& \left[ \frac{2}{s}\, F^{PC}(s)\right] 
    \left[ \alpha^2_{em}\, N_c 
    \left( \frac{4\pi \mu^2}{(s/4) \sin^2\theta_1}\right)^\epsilon 
           \frac{1}{\Gamma(1-\epsilon)}\, \frac{1}{x_1} \right]
\nonumber \\
&&\times \left( 1 + \cos^2 \theta_1 - 2\epsilon\right)
         C_F \left( \frac{\alpha_s(\mu^2)}{2\pi}\right)
        \left[ \left(\frac{4\pi\mu^2}{s}\right)^{\epsilon}
               \frac{1}{\Gamma(1-\epsilon)} \right]      
\nonumber \\
&&\times \Bigg[
  \frac{1}{\epsilon^2}
 +\frac{1}{\epsilon} 
       \left(\frac{3}{2}-\ell n\frac{\delta^2}{4}\right)
\Bigg]\delta(1-x_1) \ .
\label{r2}
\end{eqnarray}
Collecting the contributions in Eqs.~(\ref{r1v}), (\ref{r3}), and
(\ref{r2}), we find the complete next-to-leading order isolated
partonic cross section for $e^+ + e^- \rightarrow q(p_1) + X$, 
\begin{eqnarray}
E_1 \frac{d\sigma^{(1)iso}_{e^+e^-\rightarrow qX}}{d^3p_1} 
&=& 
E_1 \frac{d\sigma^{(R+V)}_{e^+e^-\rightarrow qX}}{d^3p_1} 
\nonumber \\
&=&\ \left[ \frac{2}{s}\, F^{PC} (s)\right] 
     \left[ \alpha^2_{em}\, N_c\, \frac{1}{x_1} \right]
     C_F \left( \frac{\alpha_s(\mu^2)}{2\pi} \right) 
\nonumber \\
&\times & \Bigg\{
 \left( 1 + \cos^2 \theta_1 \right) \Bigg[
 -\left(1+x_1^2\right)\left(\frac{\ell n(1-x_1)}{1-x_1}\right)_+ 
 -\frac{1+x_1^2}{(1-x_1)_+}\ell n\frac{\delta^2}{4} 
\nonumber \\
&&\quad\quad 
 +\delta(1-x_1)\left( \frac{2\pi^2}{3}-\frac{9}{2}
                     -\ell n^2\frac{\delta^2}{4} \right)
 -\frac{3}{2}\left(\frac{1}{1-x_1}\right)_+ 
 -\frac{1}{2}\left(x_1-3\right) \Bigg]
\nonumber \\ 
&& + \left(1-3\cos^2\theta_1\right) \Bigg\}
\nonumber \\ 
&-& \left[ \frac{2}{s}\, F^{PC}(s)\right] 
    \left[ \alpha^2_{em}\, N_c 
    \left( \frac{4\pi \mu^2}{(s/4) \sin^2\theta_1}\right)^\epsilon 
           \frac{1}{\Gamma(1-\epsilon)}\, \frac{1}{x_1} \right]
\nonumber \\
&&\times \left( 1 + \cos^2 \theta_1 - 2\epsilon\right)
         C_F \left( \frac{\alpha_s(\mu^2)}{2\pi}\right)
        \left[ \left(\frac{4\pi\mu^2}{s}\right)^{\epsilon}
               \frac{1}{\Gamma(1-\epsilon)} \right]      
\nonumber \\
&&\times \Bigg\{
  \frac{1}{\epsilon^2}
 +\frac{1}{\epsilon} 
       \left(\frac{3}{2}-\ell n\frac{\delta^2}{4}\right)
\Bigg\}\delta(1-x_1)\ .
\label{rv}
\end{eqnarray}
As indicated in Eq.~(\ref{e90}), $\hat{y}_{13}$ is always larger than
zero for a fixed value of $x_1\neq 1$.  Therefore, 
$E_1 {d\sigma^{(1)iso}_{e^+e^-\rightarrow qX}}/{d^3p_1}$, defined in
Eq.~(\ref{rv}), is independent of $\mu_{\overline{\rm MS}}^2$.  
Furthermore, because there are no collinear subtraction terms for the
isolated partonic cross section, the short-distance isolated partonic
cross section $\hat{\sigma}^{(1)iso}={\sigma}^{(1)iso}$ given in 
Eq.~(\ref{rv}).  We note that for $x_1\neq 1$, Eq.~(\ref{rv}) reduces
to Eq.~(\ref{e67}), as it should.  However, as $x_1\rightarrow 1$, the
uncanceled $1/\epsilon^2$ and $1/\epsilon$ poles 
in Eq.~(\ref{rv}) for $\hat{\sigma}^{(1)iso}$ signal a breakdown of
conventional perturbative QCD factorization \cite{BGQ2}.

The singularities in Eq.~(\ref{rv}) corresponding to the uncanceled
poles are infrared in nature and, as expected, are proportional to
$\delta(1-x_1)$.  As explained above, the uncanceled poles come from
the interval specified by the \underline{second} term in
Eq.~(\ref{e91}).  This interval results from the restricted phase
space for soft real gluon emission due to the isolation constraints.
These poles would be irrelevant if $x_1\neq 1$.
However, $x_1\equiv x_\gamma/z$, and $x_1=1$ is kinematically allowed
when $x_\gamma=1/(1+\epsilon_h)$.  

We conclude that the conventional factorization theorem for the cross
section of isolated photons in $e^+e^-$ annihilation necessarily
breaks down when $x_\gamma \sim 1/(1+\epsilon_h)$.
%%%%%%%%%%%%%% End of Section IV %%%%%%%%%%%%%%%%%%%%%%%%%%%%%%%%%%%%%%%

%%%%%%%%%%%%%% Begin Section V %%%%%%%%%%%%%%%%%%%%%%%%%%%%%%%%%%%%%%%%%
\section{Numerical Results and Discussion}
\label{sec:5}

In this section, we present numerical results for the isolated photon
cross section in hadronic final states of $e^+e^-$ annihilation, and
we discuss the consequences of the breakdown of conventional
perturbative factorization.

In the analysis of Section~\ref{sec:4}, three regions are identified.
In the first region, $x_\gamma < 1/(1+\epsilon_h)$, the isolated cross 
section is well behaved in perturbation theory except for an infrared
logarithmic divergence as $x_\gamma$ approaches $1/(1+\epsilon_h)$,
associated uniquely with the $O(\alpha_s)$ quark-to-photon
fragmentation component of the cross section.  The second region is
the singular point $x_\gamma=1/(1+\epsilon_h)$.  At this point, the
isolated cross section is undefined in perturbation theory.  It is the
point at which the conventional factorization theorem breaks down.
The third region is the region $x_\gamma > 1/(1+\epsilon_h)$.
In this third region, but for the incompatibility of the collinear
and cone definition of fragmentation, there should be no subtraction
term and the isolated and inclusive cross sections would be identical.
However, because of this incompatibility, the $O(\alpha_s)$
subtraction terms are finite in a very small interval
$1/(1+\epsilon_h) < x_\gamma < x_\gamma^{\rm max}$, 
and the perturbative cross section is
predicted to show a logarithmic divergence as $x_\gamma$ approaches
$1/(1+\epsilon_h)$ from above.  This logarithmic divergence has a
collinear origin.  For $x_\gamma > x_\gamma^{\rm max}$ the isolated
cross section equals the full inclusive cross section.  All of
these features are illustrated in the numerical results presented
below. 

The analytical expressions, Eqs.~(\ref{e63}), (\ref{e64}) and
(\ref{e66}), for the cross sections with 
$x_{\gamma}$ less than $1/(1+\epsilon_h)$ should be useful for
comparison with data from LEP.  The expressions for $x_\gamma^{\rm
max} > x_{\gamma} > 1/(1+\epsilon_h)$, 
Eqs.~(\ref{e70}), (\ref{e72}) and (\ref{e73})
should not be taken seriously because of the logarithm, 
$\ell n((1+\epsilon_h)-1/x_{\gamma})$, caused by the
incompatibility between the two definitions of
the parton fragmentation.  For $x_\gamma > x_\gamma^{\rm max}$, the
isolated cross section equals to the inclusive cross section.
Although the phase space above
$x_{\gamma} = 1/(1+\epsilon_h)$ is small, the behavior of the
expressions for $x_{\gamma} > 1/(1+\epsilon_h)$ has important
implications for the isolated photon cross section in hadronic
collisions.  In hadronic collisions, the influence of this limited
phase space is enhanced after the hard cross section is 
convoluted with beam hadrons' parton distributions.
 
\subsection{Numerical Results}
\label{subsec:5a}

Using the analytical results derived in the previous section, we
evaluate the cross section for isolated photons at the LEP energy
$\sqrt{s}\simeq M_Z$, and we compare this cross section with
corresponding cross section for inclusive photons.  
In deriving our numerical results, we set
$M_Z=91.187$~GeV and $\Gamma_Z=2.491$~GeV.  The vector and
axial-vector couplings and other constants are taken from
Ref.~\cite{PDB}.  The weak mixing angle $\sin^2\theta_{w}=0.2319$.
For the electromagnetic coupling strength $\alpha_{em}$, we use the
solution of the first order QED renormalization group equation, and
let $\alpha_{em}(M_Z)=1/128$ \cite{BGQ1}.  For the $O(\alpha_s)$
contributions, we employ a two-loop expression for $\alpha_s(\mu^2)$
with quark threshold effects handled properly.  
We set $\Lambda_{QCD}^{(4)}=0.231$~GeV, which corresponds to
$\alpha_s(M_Z)=0.112$.  

For the quark-to-photon and gluon-to-photon fragmentation functions,
we use the analytical expressions provided in
Ref.~\cite{JFO}.  These fragmentation functions are leading-order
functions.  In principle, it would be more consistent to use
$\overline{\rm MS}$ fragmentation functions evolved in next-to-leading
order \cite{BGQ1}.  However, since our primary purpose here is 
to provide a theoretical framework for understanding the behavior of
isolated photon production, not necessarily to present the most
up-to-date numerical predictions, we believe these leading-order
fragmentation functions are adequate for demonstrating the features of
the cross section for isolated photons.

In presenting results, we divide our cross sections by an energy
dependent cross section $\sigma_0$ that specifies the leading-order
total hadronic event rate at each value of~$\sqrt{s}$:
\begin{equation}
\sigma_0 = {{4\pi s} \over {3}} \sum_q
\left[ {2\over s} F^{PC}_q (s)\ \alpha^2_{em}(s) N_c\right].
\label{www}
\end{equation}
This procedure divides out overall multiplicative factors, and, in 
doing so, shows approximately what fraction of the total hadronic 
rate is represented by prompt photon production.  One could, of
course, divide instead by the next-to-leading-order total hadronic
event rate, differing from Eq.~(\ref{www}) only by the overall factor
$(1 + \alpha_s/\pi)$.

In Fig.~\ref{fig11}, we compare the cross sections for
isolated and inclusive photons as a function of the photon energy
$E_\gamma$ at $\sqrt{s} =$91 GeV and $\theta_\gamma=90^\circ$.  We set
the renormalization scale equal to the fragmentation scale, and both
of them equal to photon's energy.  The isolation cone size is chosen
to be $\delta=20^\circ$.  The isolation energy parameters are chosen
as $\epsilon_{h}=0.05$ and $\epsilon_h=0.15$ in
Fig.~\ref{fig11}(a) and Fig.~\ref{fig11}(b), respectively.  The value
$\epsilon_h=0.05$ is typical of LEP experiments, while   
$\epsilon_h=0.15$ is used by CDF and D0 at the Fermilab Tevatron collider.
As expected, the isolated cross section is finite and well-behaved in
most of interval of $E_{\gamma}$ less than the critical value
$\sqrt{s}/2(1+\epsilon_h)$ (or $x_{\gamma} \ll 1/(1+\epsilon_h)$. 
The perturbatively calculated isolated cross section is less than the
inclusive cross section in most of this interval, as it ought to be. 
When $E_{\gamma}$ approaches $\sqrt{s}/2(1+\epsilon_h)$ (or
$x_{\gamma} \rightarrow 1/(1+\epsilon_h)$), the perturbatively
calculated isolated cross section becomes larger than
inclusive cross section, which does not make physical sense, and
it eventually approaches infinity when
$x_{\gamma}=1/(1+\epsilon_h)$, where conventional factorization 
breaks down.  When $E_{\gamma}$ approaches the critical
value from above (i.e., $x_{\gamma} > 1/(1+\epsilon_h)$), the
perturbatively calculated isolated
cross section approaches negative infinity, a result of
the term proportional to $\ell n((1+\epsilon_h)-1/x_{\gamma})$ in
Eqs.~(\ref{e70}), (\ref{e72}) and (\ref{e73}).  

Figure~\ref{fig11}(b) shows that the
perturbatively calculated isolated cross section starts  
at negative infinity when $E_{\gamma}=\sqrt{s}/2(1+\epsilon_h)$.
It increases as $E_{\gamma}$ increases and quickly becomes equal to
the inclusive cross section.  Since the values of
$x_\gamma^{\rm max}(\delta,\epsilon_h)$ in Eq.~(\ref{e32a}) and
$x^{\rm max}(z,\delta,\epsilon_h)$ in Eq.~(\ref{e44a}) are very close
to the critical value $1/(1+\epsilon_h)$, the region in which
Eqs.~(\ref{e70}), (\ref{e72}) and (\ref{e73}) are applicable is
extremely small.  Therefore, the isolated cross section is equal to
the inclusive cross section in most of the interval of $E_\gamma$
larger than the critical value $\sqrt{s}/2(1+\epsilon_h)$ 
(or $x_{\gamma} > 1/(1+\epsilon_h))$. 
 
In Fig.~\ref{fig12}, we show the breakdown of the isolated cross
section in terms of contributions from individual pieces. 
The isolated cross section is plotted as a function of $E_\gamma$ at
$\sqrt{s} = 91$~GeV and scattering angle $\theta_\gamma=90^\circ$.  
We choose $\delta=20^\circ$, $\epsilon_h=0.15$ and 
$\epsilon_h=0.30$ in Fig.~\ref{fig12}(a) and Fig.~\ref{fig12}(b),
respectively.  We use the label ``0th-Frag'' for the lowest-order
fragmentation contribution, ``Direct'' for $O(\alpha_{em})$ direct
production, and ``q-Frag'' and ``g-Frag'' for the $O(\alpha_s)$ quark
and gluon fragmentation  contributions, respectively.  
The curves in Fig.~\ref{fig12} show that the breakdown of
factorization is associated with the $O(\alpha_s)$ quark fragmentation
contribution, which is infrared in nature.  All one-loop contributions
approach negative infinity as $E_\gamma$ approaches the critical value
from above, confirming that this divergence is of collinear origin,
caused by the incompatibility between the two definitions of
parton fragmentation: exact collinear long-distance fragmentation
and the short-distance contribution with a finite cone size.  

In Fig.~\ref{fig13}, we examine the predicted $\theta_\gamma$
dependence of the isolated cross section when $E_\gamma$ is about half
of the critical value.  We plot the isolated cross section
as a function of scattering angle $\theta_\gamma$ for $E_\gamma=20$
GeV.  Other parameters are chosen as shown in the figure.  There is no 
contribution from the leading order quark fragmentation, and the
$O(\alpha_s)$ contributions are negligible at this energy.  The 
angular dependence of the isolated cross section at $\sqrt{s}\simeq
20$~GeV is determined by the direct contribution, which has both 
$1+\cos^2\theta_\gamma$ and $\sin^2\theta_\gamma$ contributions (c.f.,
Eq.~(\ref{e63})). 

In Figs.~\ref{fig11}--\ref{fig13}, we set the
renormalization and fragmentation scale to be $\mu = E_\gamma$. 
Dependence of the isolated cross section on $\mu$ is examined in
Fig.~\ref{fig14}.  We plot the isolated cross section as a function of
$\mu/E_\gamma$ at $\sqrt{s}=91$~GeV, $\theta_\gamma=90^\circ$, and a
fixed $E_\gamma=20$~GeV.  Other parameters are as shown in the figure.
As expected, the isolated cross section has a very small scale
dependence.  Isolation eliminates most of the fragmentation
contributions, and the dominant direct contribution, Eq.~(\ref{e63}),
is not affected by strong interactions.  It has neither
renormalization nor fragmentation scale dependence.

Dependence on $\epsilon_h$ is strong in the region near 
$x_\gamma \sim 1/(1+\epsilon_h)$, but the perturbative calculation is 
not reliable in this region.  In Fig.~\ref{fig15}, we plot the
isolated cross section as a 
function of $\epsilon_h$ at a fixed value of $E_\gamma = 20$~GeV,
about half of the critical value.  Relatively
small dependence on $\epsilon_h$ is observed.  As expected,
Fig.~\ref{fig15} shows that the isolated cross section grows
as $\epsilon_h$ increases.  At larger $\epsilon_h$, more hadronic
energy is allowed in the isolation cone, and consequently, more events
contribute to the isolated cross section.

The isolated cross section is predicted to show a specific $\ell n\,
\delta^2$ dependence on the cone size $\delta$.  This prediction is
valid for small values of $\delta$ since we neglect throughout terms
proportional to finite powers of $\delta$.  In doing our analysis, we
explicitly constrain $\delta\neq 0$.  If $\delta$ is set to zero
before the phase space integrals are done (e.g., in Eqs.~(\ref{e32b}),
(\ref{e50}), and (\ref{e91})) then we would find
$\hat{\sigma}^{(1)sub}=0$ and $\hat{\sigma}^{isol} =
\hat{\sigma}^{incl}$.  For $x_\gamma < 1/(1+\epsilon_h)$, the $\ell
n\,\delta^2$ dependence of the isolated cross section is displayed
explicitly in Eqs.~(\ref{e63}), (\ref{e65}), and (\ref{e67}).

\subsection{Discussion}
\label{subsec:5b}

According to the conventional factorization theorem of perturbative
QCD \cite{AEMP}, the
one-loop fragmentation contributions derived here become the collinear 
counter-terms for the two-loop direct contribution to $e^+e^-
\rightarrow \gamma X$ at order $O(\alpha_{em}\alpha_s)$.  In similar
fashion to the derivation of Eq.~(\ref{e16}), we apply Eq.~(\ref{e8})
perturbatively to the order $\alpha_{em}\alpha_s$.  We obtain
\begin{eqnarray}
\left. \sigma^{(2)sub}_{e^+e^- \rightarrow\gamma X} 
   \right|_{E_q (or E_{\bar{q}})\geq \epsilon_h E_\gamma}
&=&\hat{\sigma}^{(2)sub}_{e^+e^- \rightarrow\gamma X} (x_\gamma)
   \dot{\otimes} D^{(0)}_{\gamma\rightarrow\gamma} (z)
 + \hat{\sigma}^{(2)incl}_{e^+e^- \rightarrow\gamma X} (x_\gamma)
   \ddot{\otimes} D^{(0)}_{\gamma\rightarrow\gamma} (z)
\nonumber \\
&+&\hat{\sigma}^{(1)sub}_{e^+e^- \rightarrow qX} (x_q)
   \dot{\otimes} D^{(1)}_{q\rightarrow\gamma} (z)
 + \hat{\sigma}^{(1)incl}_{e^+e^- \rightarrow q X} (x_q)
   \ddot{\otimes} D^{(1)}_{q\rightarrow \gamma} (z)
\nonumber \\
&+&\hat{\sigma}^{(0)sub}_{e^+e^- \rightarrow qX} (x_q)
   \dot{\otimes} D^{(2)}_{q\rightarrow\gamma} (z)
 + \hat{\sigma}^{(0)incl}_{e^+e^- \rightarrow q X} (x_q)
   \ddot{\otimes} D^{(2)}_{q\rightarrow \gamma} (z)
\nonumber \\
&+&(q \rightarrow \bar{q})\ .
\label{e102}
\end{eqnarray} 
Since the zeroth-order subtraction term 
$\hat{\sigma}^{(0)sub}_{e^+e^- \rightarrow qX}$ vanishes, and the
zeroth order photon-to-photon fragmentation function
$D^{(0)}_{\gamma\rightarrow\gamma} (z) = \delta(1-z)$, the factorized
expression for the two-loop short-distance hard-part becomes
\begin{eqnarray}
\hat{\sigma}^{(2)sub}_{e^+e^- \rightarrow \gamma X} (x_\gamma)
&=&
\left. \sigma^{(2)sub}_{e^+e^- \rightarrow\gamma X} 
   \right|_{E_q (or E_{\bar{q}})\geq \epsilon_h E_\gamma}
\nonumber \\
&-&\hat{\sigma}^{(1)sub}_{e^+e^- \rightarrow qX} (x_q)
   \dot{\otimes} D^{(1)}_{q\rightarrow\gamma} (z)
 - \hat{\sigma}^{(1)incl}_{e^+e^- \rightarrow q X} (x_q)
   \ddot{\otimes} D^{(1)}_{q\rightarrow \gamma} (z)
\nonumber \\
&-&\hat{\sigma}^{(0)incl}_{e^+e^- \rightarrow q X} (x_q)
   \ddot{\otimes} D^{(2)}_{q\rightarrow \gamma} (z)
 - (q \rightarrow \bar{q})\ .
\label{e103}
\end{eqnarray} 
It is clear from Eq.~(\ref{e103}) that the one-loop subtraction term, 
$\hat{\sigma}^{(1)sub}_{e^+e^- \rightarrow qX}$ derived in this
paper, Eqs.~(\ref{e52}) and (\ref{e54}), becomes a 
collinear counter-term for the two-loop direct contribution.  
Therefore, due to the breakdown of factorization at $x_\gamma =
1/(1+\epsilon_h)$ at one-loop order and the fact that the
perturbatively calculated isolated cross section is larger than the
corresponding inclusive cross section, 
the higher-order contributions to the isolated cross sections may be 
ill-defined as well.  To resolve this issue, it will be important
and necessary to investigate possible alternative definitions of the
cross section of isolated photons to all orders in perturbation
theory. 

In Eq.~(\ref{e2}), we define the isolated cross section as a
difference between the inclusive cross section and a subtraction term. 
The inclusive cross section is well-defined in perturbative QCD
\cite{BGQ1}.  Since the isolated cross section is an experimental
measurable and known to be finite, the subtraction term, defined
in Eq.~(\ref{e2}), should be finite as well.  It is not difficult to
show, as we have in this paper, that the subtraction term is an infrared
sensitive quantity in perturbation theory.  Therefore, the issue is
whether we can find a scheme
to consistently factorize the infrared sensitive quantity into a
perturbatively calculable infrared safe hard-part convoluted with
a well-defined long-distance universal function, such as the
fragmentation function.  

Although the breakdown of factorization is demonstrated in this paper
only at one-loop level in perturbation theory, the following intuitive
argument generalizes our thinking.  As remarked in
Sec.~\ref{sec:2}, the subtraction term can be viewed as a ``cross
section'' for a photon ``jet'' with photon momentum $\ell$ and
hadronic energy $E_h^{cone}$ in the ``jet'' cone, with the hadronic
energy required to be \underline{larger} than 
$E_{\max}=\epsilon_h E_\gamma$.  This ``jet cross section'' is
an integrated jet cross section, where ``integrated'' denotes 
the cross section for ``jet'' events with hadronic energy greater than
a fixed minimum value ($E_{\max}=\epsilon_h E_\gamma$) and up to a
whatever maximum value is allowed by the kinematics.  Experience with
the definition and calculation of jet cross section \cite{SW,ES}
suggests that the integrated ``jet cross section'' could be
finite in perturbative QCD, as long as we do not specify details
within the jet.  However, if we insist on determining the energy of
one specific parton within the ``jet'', and ask how this parton
fragments into the observed photon, we will most likely encounter an 
incomplete cancellation of infrared and collinear singularities.
Introduction of parton-to-photon fragmentation functions for the
cross section of isolated photons is similar to asking for the details
within the photon ``jet''.  Further analysis of these issues is beyond
the scope of the present paper.

\subsection{Phenomenology}
\label{subsec:5c}

The $O(\alpha_{em})$ direct contribution to the isolated cross section
is dominant over most of the region $x_\gamma < 1/(1+\epsilon_h)$
where perturbation theory is reliable.  This statement is illustrated
in Figs.~\ref{fig12}-\ref{fig15}.  Correspondingly, a direct
comparison of our expressions with data on isolated prompt photon
production is tantamount primarily to a verification of the
$O(\alpha_{em})$ 
expression presented in Eq.~(\ref{e63}).  In this expression, we
display the explicit dependence of the cross section on $E_\gamma$,
$\theta_\gamma$, and $\delta$.  On the other hand, an important goal
of investigations of $e^+e^-\rightarrow\gamma X$ is also the extraction
of quark-to-photon and gluon-to-photon fragmentation functions
\cite{LEP4,LEP5,LEP}.  Our analysis of this paper indicates that the
task is not straightforward.  Not only is the direct contribution
dominant over a large interval if $\epsilon_h$ is small, but as we
demonstrated to order $O(\alpha_s)$, the perturbatively computed
quark-to-parton fragmentation contribution to the isolated cross
section is infrared singular in the neighborhood of $x_\gamma =
1/(1+\epsilon_h)$.  This means that we cannot offer a reliable
theoretical expression for comparison with experiment in this region.
We expect that further theoretical work in the near future will lead
to a resolution of this unsatisfactory situation.  Meanwhile, we
suggest a possible procedure for analysis of the data.  We suggest
that the $O(\alpha_{em})$ direct contribution, specified in
Eq.~(\ref{e63}), be subtracted from the data.  Since the direct
contribution is of purely electromagnetic origin at $O(\alpha_{em})$,
it is not subject to hadronic uncertainties.  After the subtraction is
done, one is left with a sample of events that may be attributed to
$O(\alpha_s)$ quark-to-photon and gluon-to-photon fragmentation.  It
would be instructive to examine the $E_\gamma$, $\theta_\gamma$, and
$\delta$ dependence of this subtracted sample.

Isolation enhances the $O(\alpha_{em})$ direct contribution and
eliminates the leading-order quark-to-photon fragmentation
contribution for $x_\gamma$  
less than the critical value $1/(1+\epsilon_h)$.  
It has been suggested \cite{EWNG} that the region in which the
contribution from leading-order quark-to-photon fragmentation is
important can be enlarged if $\epsilon_h$ is chosen to be relatively
large.  In Figs.~\ref{fig12}(b) and \ref{fig12}(c), the dashed curve
shows the contribution from the leading-order fragmentation
contribution; it contributes only in the region above
$x_\gamma=1/(1+\epsilon_h)$.   
The leading-order fragmentation contribution dominates in
this region, and the isolated cross section is essentially the same as
the inclusive cross section \cite{BGQ1}.  
Experimental measurement of prompt photons in this region is difficult
both because the rate is low and because clean identification of a {\it
prompt} photon signal is problematic if $\epsilon_h$ is relatively
large.  When $\epsilon_h$ is large, the hadronic energy in the cone
makes it difficult on an event-by-event basis to be certain the photon
does not result from $\pi^o$ or $\eta$ decay, or from other hadronic
processes.  

We turn next to implications of our analysis for the computation of 
isolated prompt photon cross sections at hadron colliders.  In hadron
reactions, the lowest-order subprocesses are the two
$O(\alpha_s\alpha_{em})$ direct subprocesses, $q + g \rightarrow q +
\gamma$ and $q + \bar{q} \rightarrow g + \gamma$, along with the
lowest-order fragmentation subprocesses.  In the lowest-order
fragmentation, various $O(\alpha_s^2)$ two-parton to two-parton 
subprocesses are followed by leading-order long distance
quark-to-photon or gluon-to-photon fragmentation.   In computing
prompt photon production to next-to-leading order in QCD, one must
compute all  $O(\alpha_s^2\alpha_{em})$ direct subprocesses and
include fragmentation at next-to-leading order.  It is at this level
that problems of the type we find in our study of electron-positron
annihilation will also be encountered in the hadronic case. 
Specifically, the next-to-leading order quark fragmentation
contribution is ill defined owing to the infrared problem identified
in our analysis.  In their treatment of prompt photon production,
Baer, Ohnemus, and Owens \cite{Baer} include the direct contributions
through $O(\alpha_s^2\alpha_{em})$, but they include fragmentation
through lowest-order only.  Their results are therefore not affected
by the infrared problem.  The studies of Gordon, Vogelsang, and 
collaborators \cite{GV} include fragmentation through next-to-leading
order.   

It is not our intent in this paper to provide a new analysis of the 
hadronic case.  Rather, we wish to point out that the infrared
sensitivity of the next-to-leading order quark-to-photon fragmentation
contribution is enhanced in the hadronic case.  We also suggest a
temporary phenomenological remedy.  In hadron-hadron collisions, we
are interested in the production of prompt photons as a function of
the photon's transverse momentum, $p_T$.  Invoking factorization,
and continuing to use $\ell$ to denote the four-vector 
momentum of the photon, we write the inclusive or the isolated cross
section in $hadron(A)+hadron(B)\rightarrow \gamma X$ as 
\begin{equation}
E_\gamma \frac{d\sigma_{AB \rightarrow \gamma X}}{d^3\ell} 
= \sum_{a,b} 
\int dx_a\, \phi_{a/A}(x_a)\, dx_b\, \phi_{b/B}(x_b)\,
E_\gamma \frac{d\sigma_{ab \rightarrow \gamma X}}{d^3\ell} \ .
\label{e3st}
\end{equation}
The sum is taken over all initial partons ($q, \bar{q}, g$) in the
incident hadrons, and $\phi(x)$ are parton density distributions.  The
partonic level cross section is expressed as 
\begin{equation}
E_\gamma \frac{d\sigma_{ab \rightarrow \gamma X}}{d^3\ell} 
= \sum_{c} \int \frac{dz}{z}\, 
E_c \frac{d\hat{\sigma}_{ab \rightarrow c X}}{d^3p_c} 
\left(p_c=\frac{\ell}{z}\right)\,
\frac{D_{c\rightarrow \gamma}(z)}{z} \ ,
\label{e4st}
\end{equation}
where $c = \gamma, q, \bar{q}$, or $g$.  If the factorization theorem
holds for isolated photons, the partonic hard-parts
$\hat{\sigma}_{ab\rightarrow cX}$ in Eq.~(\ref{e4st}) should be finite
for isolated photons and should be smaller than the corresponding hard
parts for the inclusive cross section.  This is not the case for the
perturbatively computed quark fragmentation contribution at one-loop,
as we now illustrate.  

To streamline the discussion we begin by defining a parton-parton flux
$\Phi(\tau)$,
\begin{equation}
\Phi_{ab}(\tau) 
= \int_\tau^1 \frac{dx}{x}\, \phi_{a/A}(x)\, \phi_{b/B}(\tau/x) \ ,
\label{e5st}
\end{equation}
where $\tau\equiv x_a\, x_b$.
To exploit our electron-positron results most conveniently, we limit our
discussion to the contribution in Eq.(\ref{e3st}) associated with the
quark-antiquark initial state, and, further, we take only the term in
which the flavors of the initial and final quarks differ:  $q' +
\bar{q}' \rightarrow q + \bar{q} + g$, where $q$ fragments to
$\gamma$.  We point out, however, that 
there is a next-to-leading order quark fragmentation contribution
associated with almost all subprocesses.  Our illustration is 
therefore much more general than the specific example we treat.  We
further specialize to prompt photon production at zero rapidity (i.e.,
at $90^\circ$ in the hadron-hadron center-of-mass frame), and ignore
real gluon emission from the initial-state quark ($q'$) antiquark
($\bar{q}'$) and from the $s$-channel virtual gluon ($g$).  As an
expedient approximation we take equal values for the partonic
fractions, $x_a = x_b = x = \sqrt{\tau}$.  This latter approximation
allows us, in the isolated case, to use our Eqs.~(\ref{e66}) and
(\ref{e67}) directly, setting $\theta_\gamma=90^\circ$ in
Eq.~(\ref{e67}).  Likewise, in the inclusive case, we use our 
Eqs.~(\ref{e61}) and (\ref{e62}) directly, setting
$\theta_\gamma=90^\circ$ in Eq.~(\ref{e62}).  The normalization factor
differs, of course, in the hadron case.  In Eqs.~(\ref{e62}) and (\ref{e67}),
we replace  
\begin{mathletters}
\label{e6st}
\begin{eqnarray}
s &\rightarrow & \hat{s}=x_a\, x_b\, s = \tau\, s\ ;
\label{e6sta} \\
\frac{2}{s} F_{q}^{PC}(s) &\rightarrow & \frac{1}{\hat{s}^2}\ ; 
\label{e6stb} \\
\alpha_{em}^2   &\rightarrow & \alpha_s^2 \ ; 
\label{e6stc} \\
N_c\, C_F & \rightarrow & \frac{N^2-1}{N^3} \ .
\label{e6std}
\end{eqnarray}
\end{mathletters}
In the translation to the hadronic case, the variable $x_\gamma$ 
becomes $\hat{x}_T$ where $\hat{x}_T = 2 p_T/\sqrt{\hat{s}} \sim
x_T/x$ with $x_T=2p_T/\sqrt{s}$.  

The special one-loop quark fragmentation contribution
to the observed cross section takes the form 
\begin{equation}
E_\gamma \frac{d\sigma_{AB \rightarrow \gamma X}}{d^3\ell} 
\sim  
\int_{x_T^2}^1 d\tau\ \Phi_{q'\bar{q}'}(\tau)\ 
E_\gamma \frac{d\sigma_{q'\bar{q}'\rightarrow \gamma
X}}{d^3\ell}(\tau) \quad + \quad \mbox{other subprocesses} \ . 
\label{e7st}
\end{equation}
In Eq.~(\ref{e7st}), the one-loop contribution $E_\gamma
d\sigma_{q'\bar{q}'\rightarrow \gamma X}/d^3\ell(\tau)$ to inclusive
[or isolated] cross section is obtained from Eq.~(\ref{e62}) [or
Eq.~(\ref{e67})] with the replacements defined in Eq.~(\ref{e6st}).

In Fig.~\ref{fig16}, we present the one-loop quark-to-photon
fragmentation contributions to the inclusive and isolated prompt
photon cross sections as a function of the scaling variable $x_\gamma$
in $e^+e^-$ annihilation.  The region of infrared instability in the
neighborhood of $x_\gamma = 1/(1 + \epsilon_h)$ is evident, as is the
fact that the isolated cross section exceeds the inclusive in part of the 
$x_\gamma$ interval.  

In Fig.~\ref{fig17}, we show the integrand in Eq.~(\ref{e7st})
obtained after convoluting the parton flux of Eq.~(\ref{e5st}) with
the one-loop contributions in Fig.~\ref{fig16} (with the replacements
in Eq.~(\ref{e6st})).  In obtaining Fig.~\ref{fig17}, we use CTEQ3M
parton distributions \cite{CTEQ} to compute the effective parton
flux, $\Phi_{q'\bar{q}'}(\tau)$ with $q'=u$ (up quark); the
final-state quark flavor is chosen as $q=d$ (down quark).  The
features demonstrated in Fig.~\ref{fig17} are independent of the choice
of quark flavors.

The one-loop fragmentation contribution to the hadronic cross section
at a given value of transverse momentum $p_T$ is obtained by computing
the area under the curve in Fig.~\ref{fig17} from 
$x_T = 2 p_T/\sqrt{s}$ to 1.  It is evident that the convolution with
the parton flux substantially enhances the influence of the region of
infrared sensitivity.  The divergences above and below the point
$\hat{x}_T=x_T/\sqrt{\tau} = 1/(1 + \epsilon_h)$ [or
$x=\sqrt{\tau}=x_T(1+\epsilon_h)$] are integrable logarithmic
divergences, and thus they yield a finite contribution if an
integral is done over all $x$ (or all $\tau$).  However, we stress
that the perturbatively calculated one-loop partonic cross section
$E_q\, d\hat{\sigma}^{iso}_{q'\bar{q}'\rightarrow q X}$, obtained
from Eq.~(\ref{e67}), has an inverse-power divergence as
$x_q\rightarrow 1$ and has uncanceled $1/\epsilon$ poles in
dimensional regularization \cite{BGQ2}.  The divergence for 
$x_{\gamma} < 1/(1 + \epsilon_h)$ becomes a logarithmic
divergence only after convolution with a long-distance
fragmentation function.  The theory is not defined 
at the point $\hat{x}_T = 1/(1 + \epsilon_h)$, and it would be
inappropriate to take the perturbatively calculated cross section at
face value when it exceeds the inclusive cross section.   

As remarked earlier, a new theoretical definition of the isolated 
photon cross section is desirable, along the lines of a "photon jet"
definition.  It would be infrared safe to all orders in $\alpha_s$.
Presumably, the new definition would avoid parton-to-photon 
fragmentation functions.
Meanwhile, we propose a simple phenomenological remedy to deal with
infrared problem illustrated in Figs.~\ref{fig16} and \ref{fig17}.  In
calculations of isolated prompt photon cross sections in hadron-hadron 
reactions, we suggest that the perturbatively calculated 
one-loop quark-to-photon 
fragmentation contribution to the isolated cross section be used only in the
region of phase space in which it is calculated to be smaller than the
inclusive rate.  Elsewhere, the one-loop inclusive contribution should
be used as an {\it upper bound}.     
%%%%%%%%%%%%%% End of Section V %%%%%%%%%%%%%%%%%%%%%%%%%%%%%%%%%%%%%%%%

%%%%%%%%%%%%%% Begin Acknoledgement %%%%%%%%%%%%%%%%%%%%%%%%%%%%%%%%%%%%
\section*{Acknowledgements}

We thank George Sterman for helpful discussions.
X. Guo and J. Qiu are grateful for the hospitality of Argonne National 
Laboratory where a part of this work was completed.  Work in the High 
Energy Physics Division at Argonne National Laboratory is supported by 
the U.S. Department of Energy, Division of High Energy Physics, 
Contract W-31-109-ENG-38. 
The work at Iowa State University is supported in part 
by the U.S. Department of Energy under Grant Nos. 
DE-FG02-87ER40731 and DE-FG02-92ER40730.
%%%%%%%%%%%%%% End of Acknoledgement %%%%%%%%%%%%%%%%%%%%%%%%%%%%%%%%%%%

%%%%%%%%%%%%%% Begin References %%%%%%%%%%%%%%%%%%%%%%%%%%%%%%%%%%%%%%%%

%%%%%%%%%%%%%% End of References %%%%%%%%%%%%%%%%%%%%%%%%%%%%%%%%%%%%%%%

%%%%%%%%%%%%%% Begin Figure Captions %%%%%%%%%%%%%%%%%%%%%%%%%%%%%%%%%%%
\begin{figure}
\caption{Illustration of an isolation cone about the direction of
the photon's momentum.}
\label{fig0}
\end{figure}

\begin{figure}
\caption{Illustration of $e^+e^- \rightarrow \gamma X$ in an $m$ parton
state: $e^+e^- \rightarrow c X$ followed by fragmentation 
$c \rightarrow \gamma X$.}
\label{fig1}
\end{figure}

\begin{figure}
\caption{Illustration of an isolation cone containing a parton $c$
that fragments into a $\gamma$ plus hadronic energy $E_{frag}$.  In addition,
the cone includes a gluon that fragments giving hadronic energy 
$E_{parton}$.}
\label{fig2}
\end{figure}

\begin{figure}
\caption{Lowest order, $O\left( \alpha^o_{em}\alpha^o_s\right)$, photon
production through quark fragmentation.}
\label{fig3}
\end{figure}

\begin{figure}
\caption{Feynman diagrams for $e^+e^- \rightarrow \gamma q\bar{q}$.}
\label{fig4}
\end{figure}

\begin{figure}
\caption{Order $\alpha_s$ Feynman diagram for the parton level
fragmentation function $D^{(1)}_{q\rightarrow\gamma}(z)$.}
\label{fig5}
\end{figure}

\begin{figure}
\caption{Order $\alpha_s$ Feynman diagrams for 
$e^+e^- \rightarrow q\bar{q} g$ that contribute to 
$e^+e^- \rightarrow \gamma X$ via $g\rightarrow \gamma$ fragmentation.}
\label{fig6}
\end{figure}

\begin{figure}
\caption{Contributions to the $O(\alpha_s)$ cross section 
$\sigma^{(1)}_{e^+e^- \rightarrow q X}$; both real gluon emission diagrams 
$(e^+e^- \rightarrow q\bar{q}g)$ and virtual gluon exchange terms are 
drawn.}
\label{fig7}
\end{figure}

\begin{figure}
\caption{Order $\alpha_s$ Feynman diagrams for the parton level
fragmentation function $D^{(1)}_{q\rightarrow q}(z)$.}
\label{fig8}
\end{figure}

\begin{figure}
\caption{Center of mass coordinate axes of an $e^+e^-$ collision with
the $z$-axis being the direction of the observed photon.}
\label{fig9}
\end{figure}

\begin{figure}
\caption{Comparison of the normalized cross sections for isolated
and inclusive photons in $e^+e^-\rightarrow\gamma X$ as a function of
photon energy $E_\gamma$: (a) $\epsilon_h = 0.05$; (b)
$\epsilon_h=0.15$.} 
\label{fig11}
\end{figure}

\begin{figure}
\caption{
Normalized isolated photon cross section as a function of
photon energy $E_{\gamma}$: (a) $\epsilon_h=0.15$; (b)
$\epsilon_h=0.30$.  The solid line is a the sum of
contributions from all individual subprocesses.  Shown also are the
individual contributions from the separate subprocesses.  The
$O(\alpha_{em})$ direct contribution (upper dotted line) nearly
saturates the total contribution for $x_\gamma < 1/(1+\epsilon_h)$.
The dashed line for $x\geq 1/(1+\epsilon_h)$ shows the
$O(\alpha_{em}^0\alpha_s^0)$ fragmentation contribution.  The
$O(\alpha_s)$ quark-to-photon and gluon-to-photon fragmentation
contributions are illustrated with the dot-dashed and the lower dotted
lines, respectively. } 
\label{fig12}
\end{figure}

\begin{figure}
\caption{
Normalized isolated photon cross section at $E_\gamma=20$~GeV as a
function of $\theta_{\gamma}$.  The solid line is the sum of all
contributions.  The $O(\alpha_{em})$ direct contribution (dashed line)
essentially saturates the total.  The $O(\alpha_s)$ quark-to-photon
(dot-dashed) and gluon-to-photon (dotted) contributions are also
shown.  }
\label{fig13}
\end{figure}

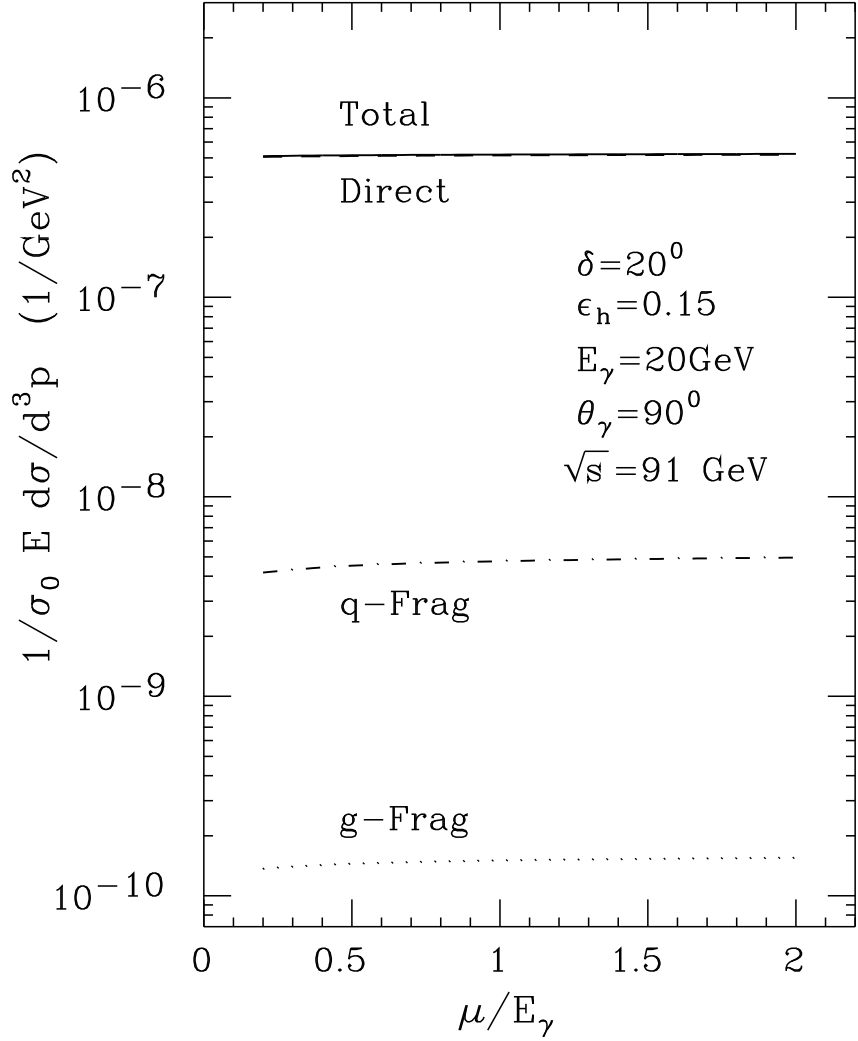
\begin{figure}
\caption{
Normalized isolated photon cross section as a function of
$\mu/E_{\gamma}$.  There is no leading order fragmentation
contribution: ``0-Frag'' at $E_{\gamma}=20$~GeV. The labeling of the
curves is the same as in Fig.~13.}
\label{fig14}
\end{figure}

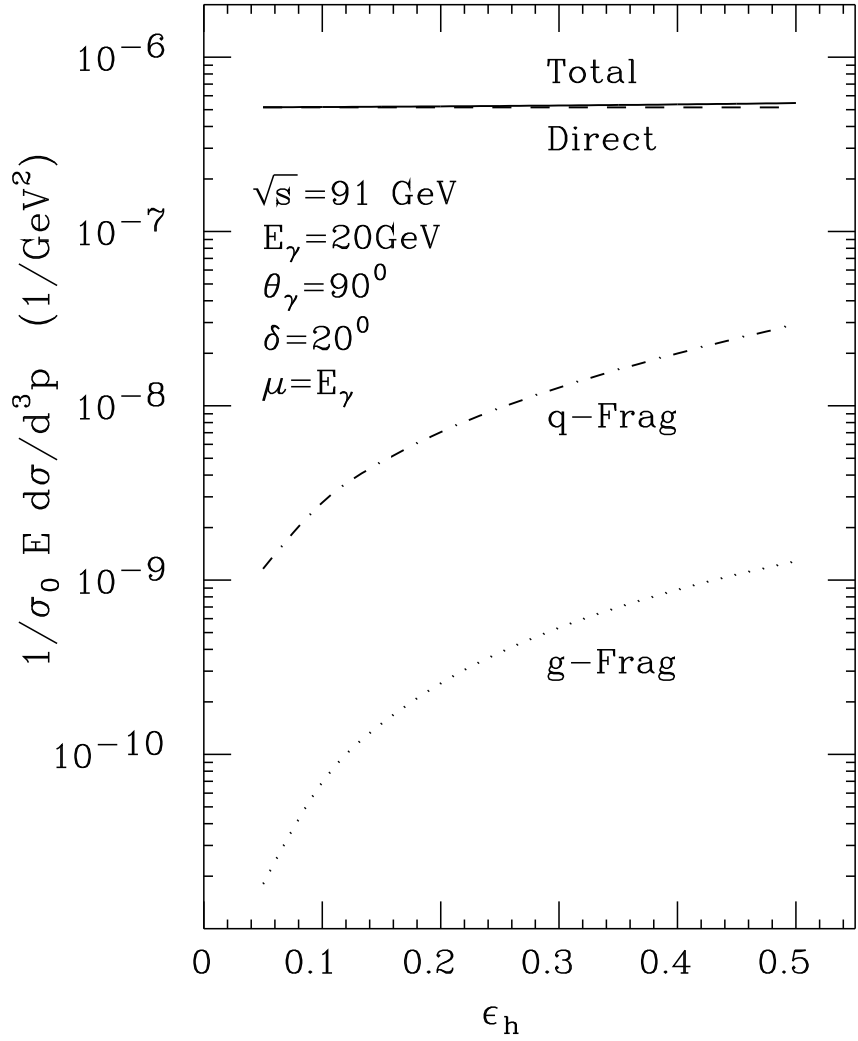
\begin{figure}
\caption{
Normalized isolated photon cross section as a function of
$\epsilon_h$.  There is no leading order fragmentation contribution:
``0-Frag'' at $E_{\gamma}=20$~GeV. The labeling of the
curves is the same as in Fig.~13.}
\label{fig15}
\end{figure}

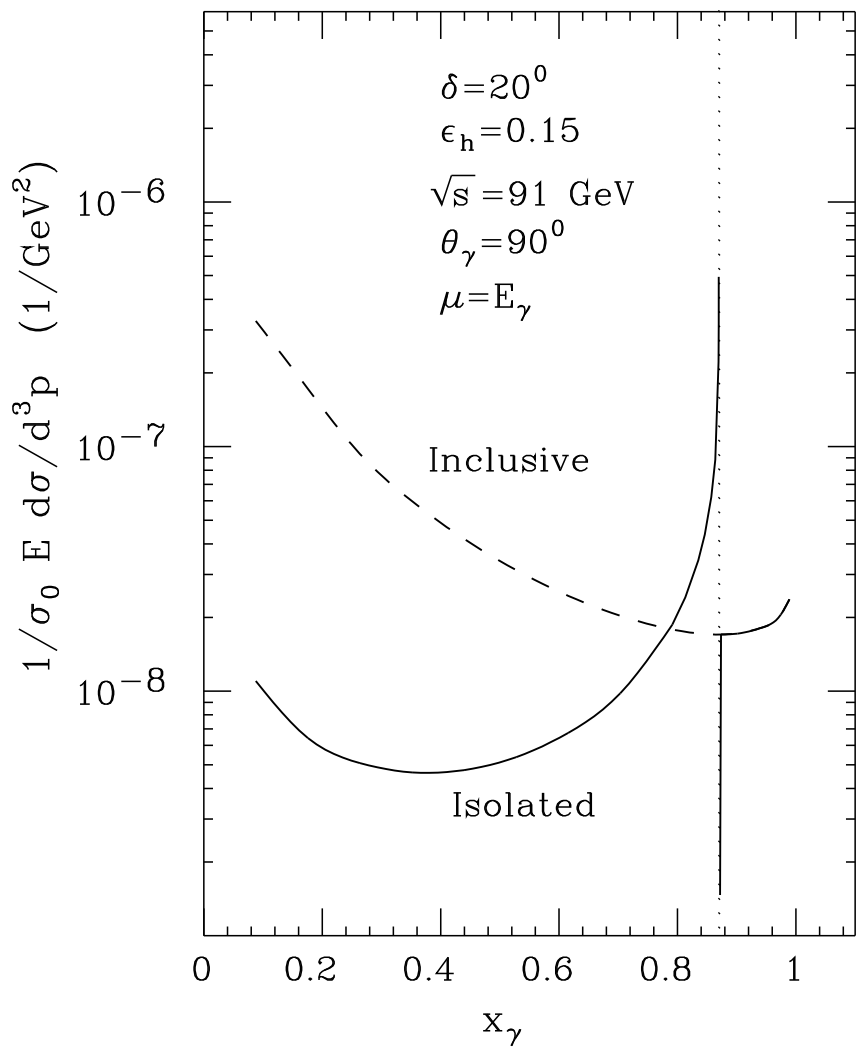
\begin{figure}
\caption{Comparison of the one-loop quark fragmentation contributions
to the isolated cross section and the inclusive cross section
in $e^+e^-\rightarrow\gamma X$ as a function of
$x_{\gamma} = 2E_{\gamma}/\protect\sqrt{s}$.}
\label{fig16} 
\end{figure}

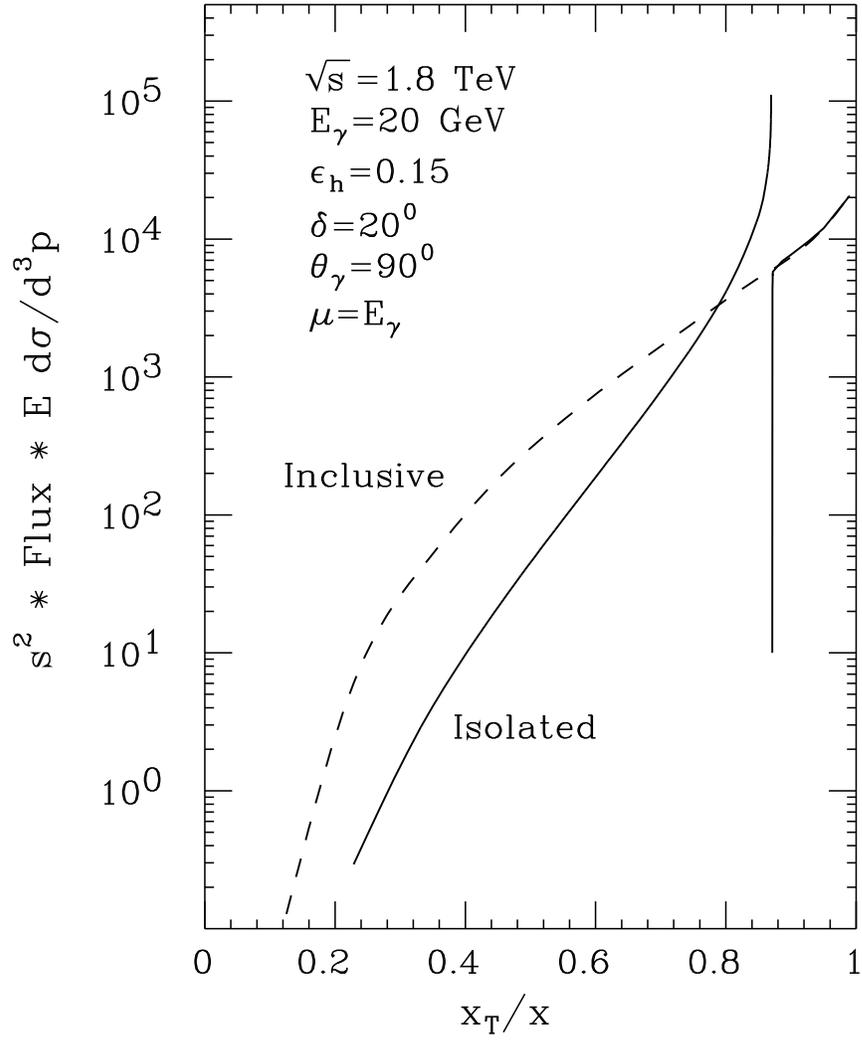
\begin{figure}
\caption{Integrand of Eq.~(110) as a function of
$x_T/\protect\sqrt{\tau}=x_T/x$ for $p \bar{p} \rightarrow \gamma X$
at $\protect\sqrt{s} =1.8$~TeV and $p_T^{\gamma} = 20$~GeV and
rapidity zero.  The isolation parameters are $\epsilon_h = 0.15$ and
$\delta = 20^\circ$.}  
\label{fig17} 
\end{figure}
%%%%%%%%%%%%%% End of Figure Captions %%%%%%%%%%%%%%%%%%%%%%%%%%%%%%%%%%

%%%%%%%%%%%%%%%%%%%%%%%%%%%%%%%%%%%%%%%%%%%%%%%%%%%

\begin{references}
\bibitem{BGQ1} 
 E. L. Berger, X. Guo, and J.-W. Qiu, Phys. Rev. {\bf D53}, 1124
(1996), and references therein. 

\bibitem{EWNG}
 E. W. N. Glover and W. J. Stirling, Phys. Lett. {\bf B295}, 128 (1992); 
 E. W. N. Glover and A.~G.~Morgan, Phys. Lett. {\bf B324}, 487 (1994).

\bibitem{KT} 
 Z. Kunszt and Z. Trocsanyi, Nucl. Phys. {\bf B394}, 139 (1993).

\bibitem{BGQ} 
 E. L. Berger, X. Guo, and J.-W. Qiu, in DPF '92, Proceedings of the 
 7th Meeting of the APS Division of Particles and Fields, Fermilab,
 November, 1992; ed. by C. H. Albright {\it{et al}}. (World
 Scientific, Singapore, 1993), Vol.~2, pp. 957-960.

\bibitem{BGQ2} 
 E. L. Berger, X. Guo, and J.-W. Qiu, Phys. Rev. Lett. {\bf 76}, 2234
(1996). 

\bibitem{BQ} 
 E. L. Berger and J.-W. Qiu, Phys. Lett. {\bf B248}, 371 (1990);
 E. L. Berger and J.-W.~Qiu, Phys. Rev. {\bf D44}, 2002 (1991).

\bibitem{Baer}
 H.~Baer, J.~Ohnemus, and J.~F.~Owens, Phys. Rev. {\bf D42}, 61
(1990).

\bibitem{Aurenche}
 P. Aurenche {\it{et al}}, Nucl. Phys. {\bf B399}, 34 (1993).

\bibitem{GV}
 L.~E.~Gordon and W.~Vogelsang, Phys. Rev. {\bf D48}, 3136 (1993)
 and {\bf D50}, 1901 (1994); 
 M.~Gl\"uck, L.~E.~Gordon, E.~Reya, and W.~Vogelsang, Phys. Rev. Lett. 
 {\bf 73}, 388 (1994).

\bibitem{AEMP} 
 G. Altarelli, R. K. Ellis, G. Martinelli, and S.~Y.~Pi, Nucl.
 Phys. {\bf B160}, 301 (1979); 
 W.~Furmanski and R.~Petronzio, Phys. Lett. {\bf B97}, 437 (1980); 
 G.~Gurci,\newline W.~Furmanski, and R.~Petronzio, Nucl. Phys.
 {\bf B175}, 27 (1980);
 J.C. Collins, D.E. Soper and G. Sterman, in {\it Perturbative Quantum
 Chromodynamics}, edited by A.H. Mueller (World Scientific, Singapore,
 1989).

\bibitem{LEP4} 
 P. D. Acton {\it{et al}}., OPAL Collaboration, Z. Phys. {\bf C58}, 
 405 (1993);
 D.~Buskulic {\it{et al}}., ALEPH Collaboration, Z. Phys. {\bf C57}, 
 17 (1993); 
 P.~Abreu {\it{et al}}., DELPHI Collaboration, Z. Phys. {\bf C53}, 
 555 (1992); 
 O.~Adriani {\it{et al}}., L3 Collaboration, Phys. Lett. {\bf B292}, 
 472 (1992).

\bibitem{LEP5} 
 D.~Buskulic {\it{et al}}., ALEPH Collaboration, Z. Phys. {\bf C69},
365 (1996). 

\bibitem{LEP} 
 P. M\"attig, H. Spiesberger, and W.~Zeuner, Z. Phys. {\bf C60}, 
 613 (1993).

\bibitem{SW} 
G. Sterman and S. Weinberg, Phys. Rev. Lett. {\bf 39}, 1436 (1977).

\bibitem{PDB} 
Particle Data Group, Phys. Rev. {\bf D50}, 1173 (1994).

\bibitem{JFO} 
J.F. Owens, Rev. Mod. Phys. {\bf 59}, 465 (1987).

\bibitem{ES} 
 S. Ellis, Z. Kunszt, and, D. Soper, Phys. Rev. Lett. {\bf 64}, 2121 
 (1990) and Phys. Rev. {\bf D46}, 192 (1992); 
 S.~D.~Ellis and D.~Soper, Phys. Rev. {\bf D48}, 3160 (1993).

\bibitem{CTEQ} 
 H.L. Lai {\it et al}., CTEQ Collaboration, Phys. Rev. {\bf D51}, 4763
(1995). 

\end{references}
\end{document}